\documentclass[aps,amsmath,amssymb,prd,showpacs,floatfix,preprint,superscriptaddress,nofootinbib,12pt]{article}

\pdfoutput=1

\usepackage{jheppub}
\usepackage{bm}
\usepackage{ulem}

\allowdisplaybreaks[3]

\begin{document}

\begin{flushright}
\small{NORDITA-2014-127}, \small{SU-ITP-14/28}
\end{flushright}

\title{Momentum dissipation and effective theories of coherent and incoherent transport}

\author[1]{Richard A. Davison}
\affiliation[1]{Lorentz Institute for Theoretical Physics, Niels Bohrweg 2, Leiden NL-2333 CA, The Netherlands}
\author[2,3,4]{and Blaise Gout\'{e}raux}
\affiliation[2]{Nordita, KTH Royal Institute of Technology and Stockholm University,
Roslagstullsbacken 23, SE-106 91 Stockholm, Sweden}
\affiliation[3]{Stanford Institute for Theoretical Physics, Department of Physics, Stanford University,
Stanford, CA 94305, USA}
\affiliation[4]{APC, Universit\'e Paris 7, CNRS, CEA, Observatoire de Paris, Sorbonne Paris Cit\'e, F-75205, Paris Cedex 13, France}
\emailAdd{davison@lorentz.leidenuniv.nl, gouterau@stanford.edu}

\abstract{We study heat transport in two systems without momentum conservation: a hydrodynamic system, and a holographic system with spatially dependent, massless scalar fields. When momentum dissipates slowly, there is a well-defined, \textit{coherent} collective excitation in the AC heat conductivity, and a crossover between sound-like and diffusive transport at small and large distance scales. When momentum dissipates quickly, there is no such excitation in the \textit{incoherent} AC heat conductivity, and diffusion dominates at all distance scales. For a critical value of the momentum dissipation rate, we compute exact expressions for the Green's functions of our holographic system due to an emergent gravitational self-duality, similar to electric/magnetic duality, and SL(2,$\mathbb{R}$) symmetries. We extend the coherent/incoherent classification to examples of charge transport in other holographic systems: probe brane theories and neutral theories with non-Maxwell actions.}

\maketitle
\clearpage
\section{Introduction}

An important feature of real-life materials is their transport properties: their response to the application of a small temperature gradient or electric field. The heat and electric currents induced by these sources are controlled by a small set of coefficients, the matrix of thermo-electric conductivities: 
\begin{equation}
\label{eq:transportmatrixintroduction}
\Bigg(\begin{array}{c}
\vec{J}\\
\vec{Q}\end{array} \Bigg) = \Bigg(
\begin{array}{cc}
\sigma & \alpha T \\
\bar\alpha T& \bar\kappa T
\end{array} \Bigg)\Bigg(\begin{array}{c}
\vec{E}\\
-\vec{\nabla}T/T\end{array} \Bigg). 
\end{equation}
At weak coupling, these coefficients can be efficiently computed using the quantum Boltzmann equations, which characterize the evolution of long-lived quasi-particle densities. At low frequencies, the AC (frequency dependent, spatially independent) conductivities will display Drude-like peaks whose height and width are set by the decay rate of the quasi-particle densities. In strongly coupled systems with no quasiparticles, different methods are needed. In systems with one ``almost-conserved'' quantity that overlaps with the electrical and heat currents, the low frequency asymptotics of the conductivity matrix exhibit Drude-like peaks characterised by the decay rate of the almost-conserved quantity, as can be shown using the memory matrix formalism \cite{forster1990hydrodynamic, Hartnoll:2007ih, Hartnoll:2012rj, Mahajan:2013cja}. One natural almost-conserved quantity is the state's total momentum, in systems where translational symmetry is only weakly broken.

When there are no almost-conserved quantities of this type in a strongly interacting system, the transport properties are more dependent upon the microscopic details of the system. In cases like this, traditional field theory calculational techniques are very complex while holographic techniques, in which transport properties are determined by studying the perturbations of higher dimensional black branes, are much simpler. Due to the strong interactions, holographic systems typically only have one possible conserved quantity that overlaps with the currents -- the state's total momentum -- which is protected in the presence of translational symmetry. When translational symmetry is weakly broken at low energies, momentum is almost-conserved, and the AC conductivities of various holographic models like this have been shown to display Drude-like peaks in agreement with memory matrix predictions \cite{Horowitz:2012ky,Horowitz:2012gs,Donos:2012js,Vegh:2013sk,Davison:2013jba,Ling:2013nxa,Blake:2013owa,Donos:2013eha,Amoretti:2014zha,Kim:2014bza,Donos:2014yya}. More generally, adapting a procedure first described in \cite{Iqbal:2008by}, DC conductivities have been computed analytically, in a variety of models where momentum is not conserved, in terms of black brane horizon data \cite{Blake:2013bqa, Blake:2013owa, Andrade:2013gsa,Davison:2013txa, Donos:2014uba, Gouteraux:2014hca, Lucas:2014zea,Blake:2014yla, Donos:2014cya, Donos:2014oha, Taylor:2014tka, Amoretti:2014mma, Donos:2014yya}. These include systems in which translational symmetry is broken strongly such that there is no almost-conserved quantity. In systems of this type, the AC and DC conductivities show departures from the Drude-like form, including transitions from conducting to insulating behaviour \cite{Donos:2012js,Donos:2013eha, Donos:2014uba, Gouteraux:2014hca,Amoretti:2014zha,Ling:2014saa, Taylor:2014tka, Mefford:2014gia, Donos:2014oha,Kim:2014bza,Baggioli:2014roa}. 

In this paper, we examine both holographic and non-holographic examples of \textit{coherent} and \textit{incoherent} conductors. Coherent conductors are those with an almost-conserved operator coupled to the associated current, such that the AC conductivity displays a Drude-like peak at low frequencies, while incoherent conductors are those where this is not the case \cite{Hartnoll:2014lpa}. We describe and explain general features present in the low energy and long distance transport properties of each of these kind of states, going beyond just the DC and AC conductivities, and also classify various holographic models in this way. We present simple, effective descriptions, typically in terms of a small number of long-lived excitations of the system, that control the low energy transport. In this paper, we focus on the simplest kind of examples: those in which the heat and electrical currents are decoupled. 

There are many real systems with anomalous transport properties in their normal state, and where varying the strength of momentum non-conservation, for instance by varying the doping, causes a crossover from a coherent to an incoherent conducting state, and ultimately to an insulator (e.g. \cite{PhysRevB.47.8233,PhysRevB.43.7942}). A better understanding of the general structure of low energy transport in each of these different kinds of strongly interacting systems may help to identify the reasons for some of these properties. 

The transport properties of a conserved charge are contained in its time- and space-resolved conductivity $\kappa$, which is proportional to the two-point retarded Green's function of its associated current. The collective excitations that carry current can be identified from the poles of the (Fourier-transformed) conductivity, after continuing the frequency $\omega$ to the complex plane. The real part of a pole frequency indicates its propagating frequency, and the imaginary part indicates its decay rate. Typically, a conductivity will have numerous poles, and only those closest to the origin of the complex $\omega$ plane determine the long-time transport properties in which we are interested. 

At an operational level, we define a coherent conductor as one in which the low frequency AC conductivity is dominated by a single, purely imaginary pole at $\omega=-i\Gamma$, which is parametrically closer to the origin than the nearest other poles at $\left|\omega\right|\sim\Lambda$, and which overwhelms any analytic contributions to the low frequency conductivity. In this case, the conductivity at low frequencies $\omega\ll\Lambda$ is well-approximated by the Drude-like result
\begin{equation}
\label{eq:drudepeakintroduction}
\kappa\left(\omega\right)=\frac{K}{\Gamma-i\omega}.
\end{equation}
The physical interpretation of this is clear: the current is carried by a long-lived collective excitation with decay rate $\Gamma$ that lives parametrically longer than the other collective excitations of the state (which decay at a rate $\sim\Lambda$). A long-lived excitation of this form will be produced by the existence of an almost-conserved operator that couples to the relevant current. In the incoherent case, the conductivity cannot be approximated by a single pole near the origin that is well-separated from the other poles at $\left|\omega\right|\sim\Lambda$. In this case, the conductivity will be approximately constant for $\omega\ll\Lambda$, where there are no collective excitations, and will depend upon the specific details of the state's excitations for $\omega\gtrsim\Lambda$. Importantly, in a system where thermal effects are important at an energy scale $\Lambda\sim T$, the transport is only coherent if the corresponding Drude-like peak has a width $\Gamma\ll T$ and so dominates the system's late time transport properties. In cases where $\Gamma$ is a tunable parameter of the system, a coherent/incoherent crossover will occur when $\Gamma\sim\Lambda$. 

We will restrict our attention to frequencies (and wavenumbers) below a UV cutoff scale $\Lambda_{UV}$ coinciding with $\Lambda$, the scale where the microscopic details of the theory become important. This allows us to extract the general features of transport i.e.~those associated with exact or approximate conservation laws, rather than those that depend upon the specific microscopic details of each theory. While an investigation of these microscopic excitations in our simple holographic theories would be worthwhile, it is not obvious that these will bear any resemblance to the excitations present in real systems.

As a simple example, consider a neutral, conformal hydrodynamic state, whose heat current $\vec{Q}$ is proportional to its momentum density $\vec{P}$. In this hydrodynamic system, the momentum is exactly conserved and thus the heat conductivity has the form (\ref{eq:drudepeakintroduction}) with $\Gamma=0$. Using the Kramers-Kr\"onig relation, this implies that the system is a perfect heat conductor: heat cannot dissipate since it carries momentum, which is conserved. Upon increasing the rate of momentum dissipation $\Gamma$, one will find a Drude-like peak of the form (\ref{eq:drudepeakintroduction}) until $\Gamma\sim\Lambda\sim T$. At this point, the momentum dissipation rate becomes of the same order as the characteristic energy scale of the system $T$ and the transport becomes incoherent. At low frequencies $\omega\ll\Lambda\sim T$ in the incoherent regime, the AC conductivity will be approximately constant, as there are no collective excitations that carry heat over such long timescales.

Further information about the system's transport properties are contained in the spatially resolved longitudinal conductivity: the response of the current to a source with spatial modulation in the direction of the current. In the simple case of homogeneous systems, it is convenient to perform a Fourier transform and work with non-zero wavenumbers $k$. When an almost-conserved operator exists that overlaps with the current (i.e. for a coherent conductor), the qualitative features of the spatially resolved conductivity will depend upon the distance scale (or $k$) at which one looks. Although there is a collective excitation that decays parametrically slower than the other excitations of the system $\Gamma\ll\Lambda$, it still has a non-vanishing decay rate $\Gamma$. Therefore, at very long distances $k\ll\Gamma\ll\Lambda$, the lowest energy excitations that carry current are those of a system \textit{without} the associated conservation law. These excitations are typically diffusive, as this is a generic feature of systems which do not have any conservation laws except those of energy and/or electric charge. It is only at shorter distance scales $\Gamma\ll k\ll\Lambda$ that the approximate conservation law is recovered, and qualitatively different low energy excitations related to this conservation law exist in this regime. In an incoherent state \cite{Hartnoll:2014lpa}, there is never an almost-conserved operator that couples to the current and so no crossover of this type occurs. Despite the absence of any features in the low frequency AC conductivity in an incoherent state, the spatially resolved conductivity is less trivial: one will typically find a diffusive excitation at all distance scales ($k\ll\Lambda$), due to the conservation law of the appropriate charge.

In our simple example of a neutral, hydrodynamic state, this means that heat diffuses at long distances in either the coherent or incoherent regimes. In the coherent case, one finds a qualitative change in behaviour at shorter distances, where heat is transported via sound-like waves due to the approximate validity of momentum conservation over these distance scales.

In this paper, we examine in detail heat transport in a neutral, relativistic hydrodynamic system, modified to incorporate momentum dissipation,\footnote{See also \cite{Balasubramanian:2013yqa} for related work.} and verify the statements given above. Firstly, we perform a field theoretic analysis, and derive the heat conductivity and excitation spectrum just outlined. We then compare these results to an explicit holographic theory, in which a hydrodynamic state (dual to the AdS$_4$-Schwarzschild black brane) is modified to include momentum dissipation by coupling it to spatially dependent sources for scalar operators (dual to massless scalar fields in the gravitational theory) that explicitly break translational symmetry \cite{Andrade:2013gsa} (a model which is similar to the Q-lattices of \cite{Donos:2013eha}). This is a useful toy model which captures the essential physics, and for which the background remains isotropic, homogeneous and analytically tractable. In particular, it admits an analytical black brane solution with one tunable dimensionless parameter \cite{Bardoux:2012tr,Bardoux:2012aw}: the ratio of the translational symmetry breaking parameter $m$ to the temperature of the black brane $T$. The parameter $m$ acts as an effective mass for some of the metric fluctuations in the gravitational theory, and, for small $m$, is proportional to the momentum dissipation rate $\Gamma\sim m^2/T$ \cite{Davison:2013jba}. For small $\Gamma\ll T$ (i.e.~$m\ll T$), this theory behaves just like a coherent hydrodynamic state with slow momentum dissipation. When $\Gamma\sim T$ ($m\sim T$), the state becomes incoherent as the AC conductivity is no longer dominated by a single long-lived pole near the origin of the complex $\omega$ plane.

When $m/T=\sqrt{8}\pi$, where the energy density of the state vanishes, there is an intriguing symmetry enhancement in the holographic theory that allows us to compute all of its two-point retarded Green's functions exactly. The gravitational degrees of freedom exhibit self-duality (similar to that of a Maxwell field in AdS$_4$ \cite{Herzog:2007ij}) such that the longitudinal momentum correlator of the dual field theory is inversely proportional to the transverse momentum correlator, and the AC heat conductivity $\kappa\left(\omega\right)$ is a constant. A further SL(2,$\mathbb{R}$)$\times$SL(2,$\mathbb{R}$) symmetry is present at this point, allowing us to determine the full spatially dependent heat conductivity $\kappa\left(\omega,k\right)$ exactly. 

Finally, we revisit some well-studied holographic systems and classify their charge transport as coherent or incoherent, based on the general principles above. In the neutral state dual to the Maxwell action in a Schwarzschild-AdS$_4$ black brane background, charge transport is incoherent. There are two simple ways to deform this theory: by turning on a charge density described by the probe DBI action (i.e.~probe brane systems), or by turning on higher derivative couplings to the Maxwell gauge field. Surprisingly, we find that it is possible to obtain coherent transport in systems like this, despite the lack of any obvious conservation law that should explain why they become perfect metals in the relevant limit.

The plan of the paper is as follows. Section \ref{section:HeatTransportGen} is devoted to the hydrodynamic analysis. In section \ref{section:AxionTheory}, we present the holographic model and its heat transport properties, before describing the special features of this model at the self-dual point $m=\sqrt{8}\pi T$ in section \ref{section:SelfDuality}. We discuss the transport of electric charge in probe brane theories and neutral states with higher-derivative couplings in section \ref{section:NonMaxwell} before concluding in section \ref{section:Discussion}. Appendices \ref{sec:numericalappendix} and \ref{sec:selfdualappendix} contain some details of our holographic Green's function calculations.

\section{Heat transport in neutral, conformal plasmas \label{section:HeatTransportGen}}

We begin by discussing one of the simplest examples of transport in a collective state: the transport of heat (or, equivalently, energy) in a neutral, conformal plasma. We are interested in the late time (and long distance) transport properties of the state i.e.~the heat conductivity $\kappa\left(\omega,k\right)$ at frequencies $\omega$ and wavenumbers $k$ much smaller than the characteristic microscopic dissipation rate of the system $\Lambda$. A conformal plasma has a conserved total energy and momentum, and is consequently described by the laws of relativistic hydrodynamics when $\omega,k\ll T$. The characteristic energy scale of the system is therefore $\Lambda\sim T$ and we will also take this to be the UV cutoff scale $\Lambda_{UV}\sim\Lambda$ on processes of interest to us.

\subsection{Slow momentum dissipation: the coherent regime  \label{section:coherent}}

The equations of motion of neutral, conformal hydrodynamics are the conservation laws for energy density $\varepsilon\equiv T^{tt}$ and momentum density $P_i\equiv T^{t{x^i}}$:
\begin{align}
&\partial_t\varepsilon+\partial_iP_i=0, \label{eq:energydensityconservation}\\
&\partial_t P_i+\partial_j\Pi_{ij}=0 \label{eq:momentumconservation},
\end{align}
where $\Pi_{ij}$ is the stress tensor. For definiteness we will consider a (2+1)-dimensional hydrodynamic system and thus the indices $i,j$ run over the two spatial directions $x,y$. These conservation equations are supplemented by constitutive relations. For small fluctuations around an equilibrium state at temperature $T$, these constitutive relations allow us to express the small fluctuations of the stress tensor $\delta\Pi_{ij}$ as
\begin{equation}
\label{eq:hydroconstrelations}
\delta\Pi_{ij}=\delta_{ij}\frac{\partial p}{\partial\epsilon}\delta\varepsilon-\frac{\eta}{\epsilon+p}\left(\partial_i\delta P_j+\partial_j\delta P_i-\delta_{ij}\partial_kP_k\right),
\end{equation}
to leading order in spatial derivatives, where $\epsilon$ and $p$ are the energy density and pressure of the equilbrium state, and $\eta$ is its shear viscosity. We can therefore write a closed set of linear equations for the time evolution of small fluctuations of the conserved charge densities $\delta\varepsilon$ and $\delta P_i$ as follows
\begin{align}
&\partial_t\delta\varepsilon+\partial_i\delta P_i=0, \label{eq:energyconservationfluctuation}\\
&\partial_t \delta P_i+\frac{\partial p}{\partial\epsilon}\partial_i\delta\varepsilon-\frac{\eta}{\epsilon+p}\partial_j\partial_j\delta P_i=0.
\end{align}

Temperature fluctuations $\delta T(\vec{k})$ around an equilibrium state source energy density fluctuations $\delta\varepsilon(\vec{k})$ due to the exponent $-\varepsilon/T$ in the Euclidean path integral. Because of the coupling of $\delta\varepsilon$ to $\delta \vec{P}$ in equation (\ref{eq:energyconservationfluctuation}), a temperature gradient causes momentum to flow in the direction of the gradient, and the heat current $\vec{Q}$ of this neutral system is therefore $\vec{Q}=\vec{P}$. The associated heat conductivity $\kappa$ is defined as 
\begin{equation}
\label{eq:thermalconductivitydefinition}
\vec{Q}=-\kappa\vec{\nabla}T,
\end{equation}
and should be positive in a stable system.\footnote{$\kappa$ is equivalent to $\bar{\kappa}$ of equation (\ref{eq:transportmatrixintroduction}) in a neutral system.} In a neutral, hydrodynamic plasma, the DC heat conductivity $\kappa_{DC}\equiv\kappa(\omega\rightarrow0,\vec{k}=0)$ is infinite: the heat current does not dissipate as it overlaps with the total momentum, which is conserved. 
 
To obtain a non-trivial DC heat conductivity, momentum (and therefore heat) must be able to dissipate. To achieve this in a simple way, we modify the momentum conservation equation (\ref{eq:momentumconservation}) of our hydrodynamic model to \cite{Hartnoll:2007ih,Davison:2013jba}
\begin{equation}
\partial_t P_i+\partial_j\Pi_{ij}=-\Gamma P_i,
\end{equation}
so that momentum now dissipates isotropically at a constant rate $\Gamma$, but does not change the functional form of the constitutive relations (\ref{eq:hydroconstrelations}). At a microscopic level, momentum dissipation could be achieved by breaking the translational symmetry of the system with a lattice or impurities, and this would also invalidate the constitutive relations of the theory. However, for sufficiently small $\Gamma$ (i.e.~for sufficiently slow momentum dissipation) the modification above is the most important one as it modifies a conservation law of the system. We will discuss shortly the limitations on the magnitude of $\Gamma$ in this description. For small fluctuations around equilibrium, the momentum conservation equation is therefore modified to 
\begin{equation}
\label{eq:modifiedmomentumconservationfluctuation}
\partial_t \delta P_i+\Gamma\delta P_i+\frac{\partial p}{\partial\epsilon}\partial_i\delta\varepsilon-\frac{\eta}{\epsilon+p}\partial_j\partial_j\delta P_i=0.
\end{equation}

With the time evolution equations (\ref{eq:energyconservationfluctuation}) and (\ref{eq:modifiedmomentumconservationfluctuation}) for the conserved charges, and knowledge of what the external sources are for each of the conserved charges, one can calculate the retarded two-point Green's functions of these charges by following the approach of Kadanoff and Martin \cite{1963AnPhy..24..419K} (see, for example, \cite{Kovtun:2012rj} for a recent description of this approach). The two-point functions of the components of $\vec{P}$ parallel and transverse to the wavenumber $k$ are, respectively,
\begin{align}
&G^R_{P^xP^x}(\omega,k)=\frac{\left(\epsilon+p\right)\left[k^2\frac{\partial p}{\partial\epsilon}-i\omega\left(\Gamma+\frac{k^2\eta}{\epsilon+p}\right)\right]}{i\omega\left(-i\omega+\Gamma+\frac{k^2\eta}{\epsilon+p}\right)-k^2\frac{\partial p}{\partial\epsilon}},\\
&G^R_{P^yP^y}(\omega,k)=\frac{-\left(\epsilon+p\right)\left(\Gamma+\frac{k^2\eta}{\epsilon+p}\right)}{-i\omega+\Gamma+\frac{k^2\eta}{\epsilon+p}}, \label{eq:transverseGreensfunctioninhydro}
\end{align}
where we have chosen the wavenumber $k$ to point in the $x$-direction, without loss of generality. For conciseness, we do not list the Green's functions of $\varepsilon$ here. Up to contact terms, they can easily be obtained from those above using the Ward identity for energy conservation.

We can determine the heat conductivity $\kappa(\omega,k)$ from the longitudinal Greens functions using the Kubo formula \cite{Hartnoll:2007ih}
\begin{align}
\label{eq:hydroheatconductivity}
\kappa(\omega,k)=\frac{i}{\omega T}\left[G^R_{P^xP^x}(\omega,k)-G^R_{P^xP^x}(0,k)\right]=\frac{i\omega s}{i\omega\left(-i\omega+\Gamma+k^2\frac{\eta}{\epsilon+p}\right)-k^2\frac{\partial p}{\partial\epsilon}},
\end{align}
where $s$ is the entropy density of the equilibrium state and we have used the relation $\epsilon+p=sT$. We pause here to emphasise that it is important that one carefully subtracts the appropriate zero frequency Green's function to obtain $\kappa$, as explained in \cite{Hartnoll:2007ih,Kovtun:2012rj}. The subtracted quantity is not zero and, if ignored, one will get the wrong answer for the imaginary part of $\kappa$.\footnote{In \cite{Amoretti:2014zha}, surprising answers were obtained for $\text{Im}(\kappa)$ in a holographic theory where this subtraction was not performed. To remedy this, the authors introduced new gravitational counterterms. It would be worthwhile to check whether the correct subtraction removes the need for these counterterms.} Although it naively looks as though this subtraction will produce $\kappa_{DC}=0$, this is not the case as the $\omega\rightarrow0$ and $k\rightarrow0$ limits do not commute. Physically, one cannot just set $k=0$ at the start because it is the non-uniformities in the source $\delta T$ that generate a heat current. In this paper, we are primarily interested in the real (dissipative) part of $\kappa$, which this subtraction does not affect.

The heat conductivity (\ref{eq:hydroheatconductivity}) tells us how heat is transported in the system. To get some intuition, consider first the limit $\Gamma=0$, in which momentum is perfectly conserved. In this limit, the AC conductivity $\kappa(\omega)\equiv\kappa(\omega,0)$ consists of a pole at zero frequency with weight $s$. The Kramers-Kr\"onig relation then implies that $\text{Re}[\kappa(\omega)]\sim s\delta(\omega)$. This is as expected: the system is a perfect heat conductor because $\vec{Q}=\vec{P}$. At non-zero wavenumbers, the $\omega=0$ pole in the AC conductivity is resolved into poles in $\kappa(\omega,k)$ representing the existence of collective modes with dispersion relations
\begin{equation}
\label{eq:hydrosounddispersionrelation}
\omega=\pm\sqrt{\frac{\partial p}{\partial\epsilon}}k-i\frac{\eta}{2\left(\epsilon+p\right)}k^2+\ldots,
\end{equation}
in the system. These are the gapless hydrodynamic sound modes, characteristic excitations of a system with a conserved momentum.

\begin{figure}
\begin{tabular}{cc}
\includegraphics[width=.45\textwidth,trim= 0 200 0 0, clip]{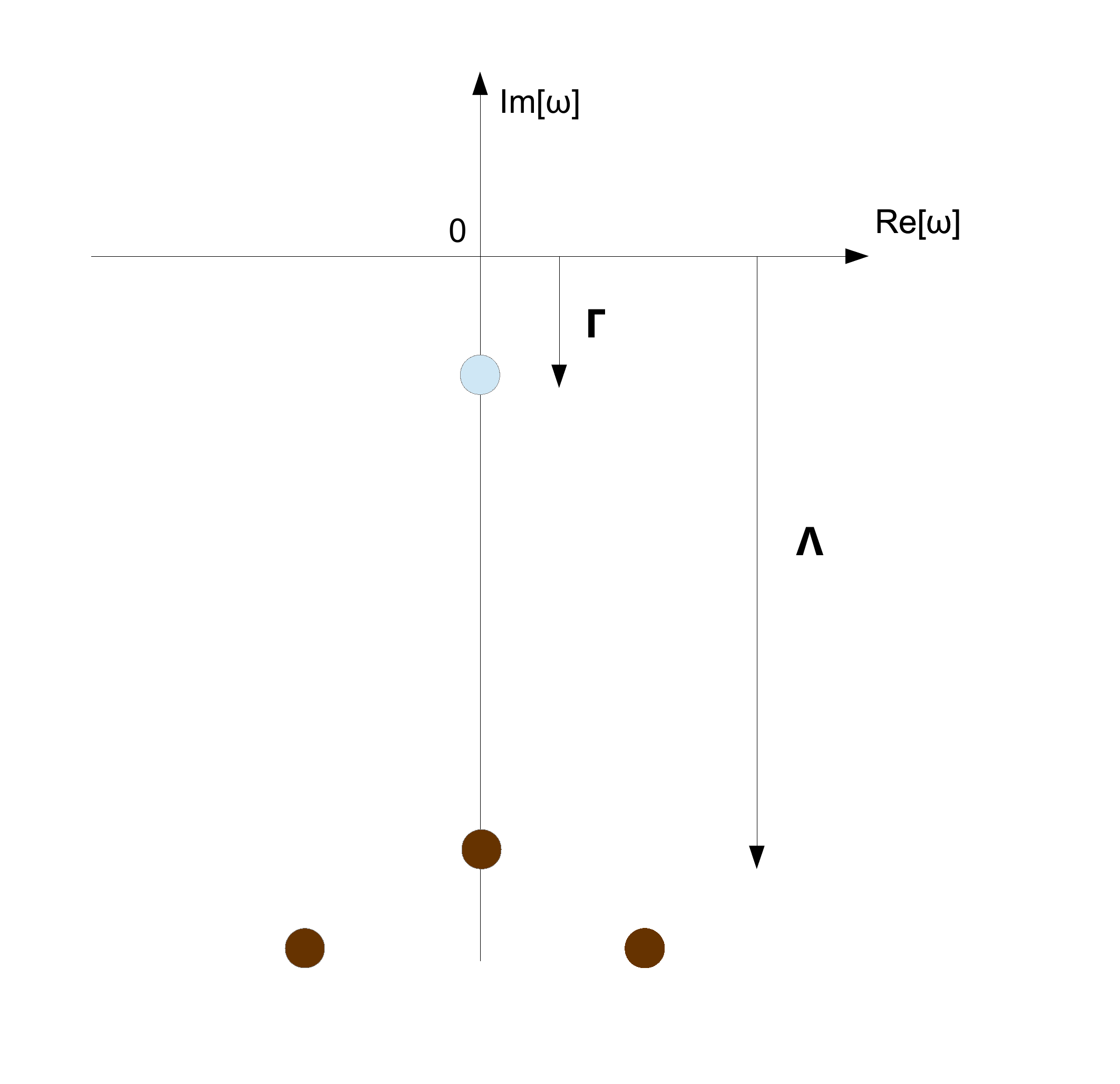}&
\includegraphics[width=.45\textwidth,trim= 0 200 0 0, clip]{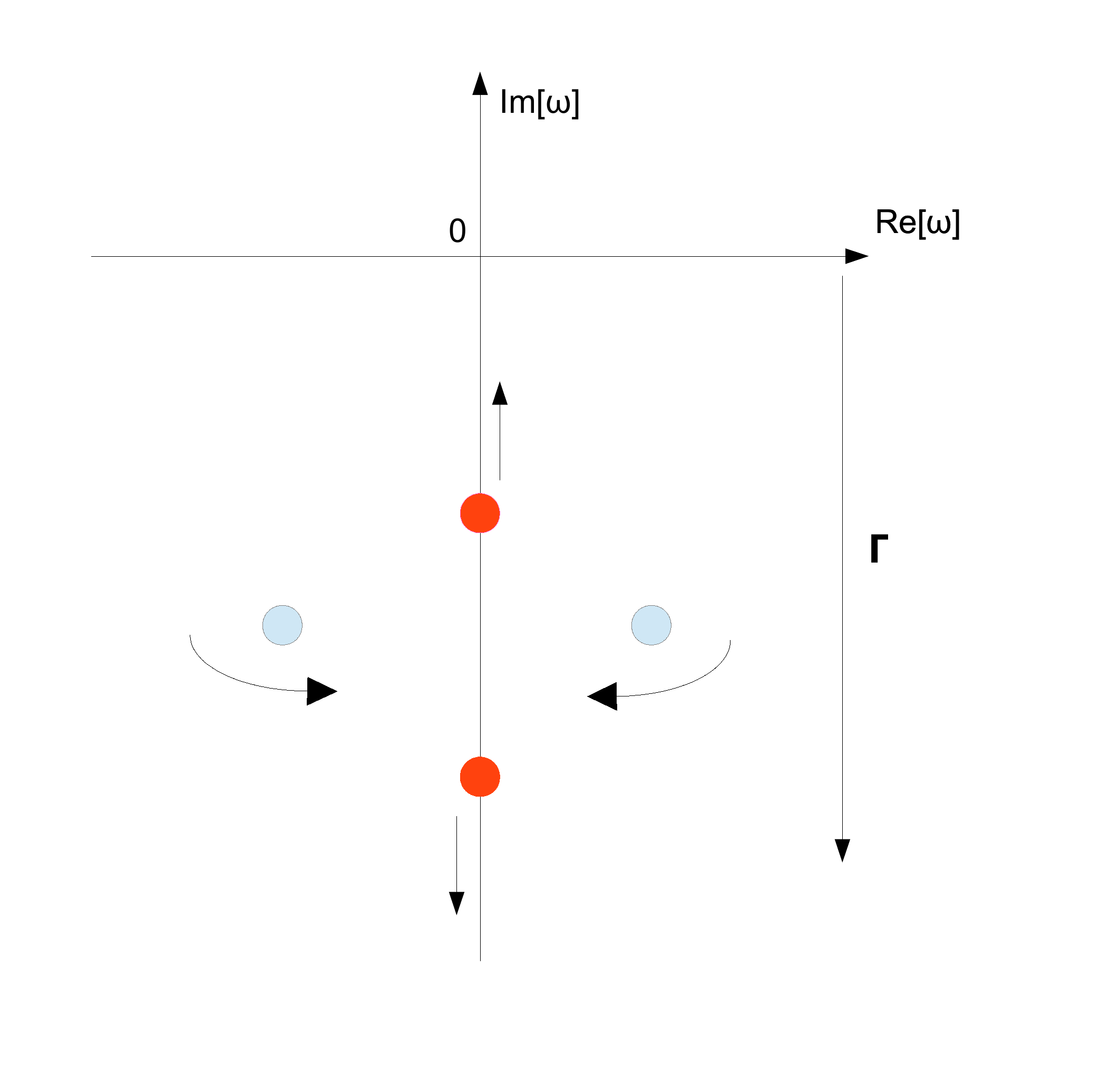}
\end{tabular}
\caption{Left: Schematic depiction of the poles of the AC heat conductivity, showing the parametric separation between the Drude-like pole (in light blue) with $\text{Im}(\omega)\sim \Gamma$ and the other poles at $\text{Im}(\omega)\sim\Lambda$ (in brown). Right: Schematic depiction of the motion of poles of $\kappa(\omega,k)$ as $\Gamma$ is increased (at fixed $k$) in the coherent regime $\Gamma\ll\Lambda$. Two sound poles (in light blue) exist when $\Gamma=0$. As $\Gamma\ll k$ is increased, these sound-like poles move deeper in the lower half plane and start getting closer to the imaginary axis. When $\Gamma\sim k$, they collide and produce two purely imaginary poles (shown in red), one of which is diffusive and moves up the imaginary axis as $\Gamma$ is increased further.}
\label{fig:PicturesQNMS}
\end{figure}

Now consider the case in which momentum dissipates at a sufficiently small non-zero rate $\Gamma$. From equation (\ref{eq:hydroheatconductivity}), we find that the zero frequency pole in the AC conductivity $\kappa(\omega)$ is resolved into a purely imaginary Drude-like pole at $\omega=-i\Gamma$:
\begin{equation}
\label{eq:coherenthydroopticalconductivity}
\kappa(\omega)=\frac{s}{-i\omega+\Gamma},
\end{equation}
the DC conductivity is finite
\begin{equation}
\label{eq:coherenthydrodcheatconductivity}
\kappa_{DC}=\frac{s}{\Gamma},
\end{equation} 
and the delta function in the real part of the AC conductivity is broadened out to a Drude-like peak of the form (\ref{eq:drudepeakintroduction}). When these equations are valid, there is a coherent response in the system: the low energy heat conductivity is controlled by the dissipation rate $\Gamma$ of a single long-lived quantity, the total momentum. However, these equations can only be trusted when the momentum dissipates slowly enough i.e.~when $\Gamma$ is small enough. The hydrodynamic description of a system is an approximation that retains only the longest-lived excitations. Going beyond this approximation, a real system of this type will have shorter-lived excitations with a typical decay rate $\Lambda\sim T$ that depend upon the microscopic physics. This is most easily visualised by examining the poles of the AC conductivity in the complex frequency plane (shown schematically in figure \ref{fig:PicturesQNMS}). When $\Gamma\ll\Lambda$, there is a single, long-lived Drude-like pole at $\omega\sim-i\Gamma$, well-separated from the microscopic excitations which produce poles at $\omega\sim-i\Lambda$ and which can therefore be neglected. This is the coherent regime. But when $\Gamma\sim\Lambda$ this is no longer possible, the total momentum no longer lives for a parametrically long time compared to the other excitations, and the system is in an incoherent regime. Furthermore, when the dissipation rate becomes $\Gamma\sim\Lambda\sim T$, the constitutive relations (\ref{eq:hydroconstrelations}) will be significantly modified due to $\Gamma$.

For now, we focus on the coherent case $\Gamma\ll\Lambda$, where momentum dissipates slowly and we can approximate the AC conductivity by the Drude-like pole (\ref{eq:coherenthydroopticalconductivity}). What are the non-zero wavenumber excitations in the theory? Recall that when $\Gamma=0$, the zero frequency pole associated with momentum conservation is resolved into sound modes. When $\Gamma\ne0$, the poles of $\kappa(\omega,k)$, from equation (\ref{eq:hydroheatconductivity}), correspond to collective excitations with dispersion relations
\begin{equation}
\label{eq:hydrocrossoverpolesanalytic}
\omega=\pm k\sqrt{\frac{\partial p}{\partial\epsilon}-\frac{1}{4}\left(\Gamma k^{-1}+\frac{\eta}{\epsilon+p}k\right)^2}-\frac{i}{2}\left(\Gamma+\frac{\eta}{\epsilon+p}k^2\right)+\ldots,
\end{equation}
where the ellipsis denotes higher order terms in the wavenumber expansion. The types of excitation described by these dispersion relations, and the physical reasons for them, depend upon whether one looks at small wavenumbers $k\ll\Gamma\ll\Lambda$ (i.e.~at distances much longer than the momentum lifetime) or at large wavenumbers $\Gamma\ll k\ll\Lambda$ (i.e.~at distances much shorter than the momentum lifetime). The case $\Gamma\ll\Lambda\ll k$ corresponds to high energy physics ($\Lambda_{UV}\ll k$), outside our region of interest.

It's easiest to first consider the case of large wavenumbers $\Gamma\ll k\ll\Lambda$. This is a small deformation of the case $\Gamma=0$, in which we previously saw that the low energy excitations are sound modes with dispersion relations (\ref{eq:hydrosounddispersionrelation}). From equation (\ref{eq:hydrocrossoverpolesanalytic}), we find that when $\Gamma/k\ll1$, the excitations have dispersion relations
\begin{equation}
\label{eq:hydrodispersionsofsoundlikemodes}
\omega=\pm\sqrt{\frac{\partial p}{\partial\epsilon}}k-\frac{i}{2}\left(\Gamma+\frac{\eta}{\epsilon+p}k^2\right)+\ldots.
\end{equation}
These are sound-like modes, as can be seen by a comparison with equation (\ref{eq:hydrosounddispersionrelation}), and the effect of the small momentum dissipation rate $\Gamma$ is to increase the decay rate of the excitation by $\Gamma$ such that it has a non-zero decay rate even at $k=0$. In this range of parameters, the sound-like mode is still a well-defined, long-lived excitation. Physically, this makes sense: wavenumbers $\Gamma\ll k$ correspond to short distances with respect to the lifetime of the momentum. Over such short distances, momentum is approximately conserved and sound-like excitations survive.

This is no longer true for excitations of very small wavenumber $k\ll\Gamma\ll\Lambda$. These excitations live at long distances with respect to the momentum dissipation time, and, although momentum dissipates slowly with respect to the microscopic excitations of the system ($\Gamma\ll\Lambda$), at these long distances momentum dissipates significantly and so the sound-like mode is destroyed. It is replaced by two purely imaginary modes, as can be seen by taking the appropriate limit $\Gamma/k\gg1$ of equation (\ref{eq:hydrocrossoverpolesanalytic}): a diffusive excitation with dispersion relation
\begin{equation}
\label{eq:hydrodiffusiondispersion}
\omega=-i\frac{\partial p}{\partial\epsilon}\Gamma^{-1}k^2+\ldots,
\end{equation}
and a Drude-like excitation
\begin{equation}
\label{eq:hydroDrudetypemodefinitek}
\omega=-i\Gamma+ik^2\left(\frac{\partial p}{\partial\epsilon}\Gamma^{-1}-\frac{\eta}{\epsilon+p}\right)+\ldots.
\end{equation}
Thus, introducing even a very small momentum dissipation rate $\Gamma$ qualitatively changes the long distance physics of the state: heat is now primarily conducted by diffusive processes at these distances scales, rather than by sound waves. No matter how slowly the momentum dissipates, at long enough distances it has dissipated significantly and diffusion takes over. At these long distances, the low energy physics is dominated by the only conservation law left in the state: energy conservation. The low energy excitation characteristic of a state in which only energy is conserved is diffusion (see, for example, the discussion in subsection \ref{section:incoherent}). The diffusion constant in (\ref{eq:hydrodiffusiondispersion}) obeys the appropriate Einstein relation (\ref{eq:incoherenthydroDCconductivity}).

As we have just described, in this coherent state heat diffuses at long distances ($k\ll\Gamma\ll\Lambda$) and propagates via sound-like waves at short distances ($\Gamma\ll k\ll\Lambda$). At intermediate distances $\Gamma\sim k$ there is a crossover between these two kinds of behaviour. It is instructive to see how this crossover manifests itself in the motion of the poles of the heat conductivity $\kappa(\omega,k)$ in the complex frequency plane. Consider first the case of increasing the wavenumber $k$ from zero, at a fixed, non-zero $\Gamma$. When $k=0$, there is only one pole near the origin: the purely imaginary Drude-like pole at $\omega=-i\Gamma$ as previously described. At small $k\ll\Gamma\ll\Lambda$, this Drude-like pole begins to move towards the origin with dispersion relation (\ref{eq:hydroDrudetypemodefinitek})\footnote{\label{footnote3}In the coherent regime, $\Gamma\ll T$ and thus $\partial p/\partial\epsilon\sim1$ and $\eta/\left(\epsilon+p\right)\sim T^{-1}$ will be close to their $\Gamma=0$ values. The $k^2$ term in the dispersion relation (\ref{eq:hydroDrudetypemodefinitek}) is therefore always positive, and so the mode moves up the imaginary axis as $k$ increases.} while a new, diffusive pole emanates from the origin and begins to move down the imaginary axis with dispersion relation (\ref{eq:hydrodiffusiondispersion}). We emphasise here that, when $k=0$, this diffusive pole does \textit{not} turn into a pole at the origin, but simply ceases being a pole.\footnote{While this may appear odd at first glance, it is perfectly normal. Consider the electrical conductivity $\sigma(\omega,k)$ in a neutral, hydrodynamic theory. Due to charge conservation, the conductivity is governed by diffusion $\sigma(\omega,k)=i\omega\chi D/(i\omega-Dk^2)$. For non-zero $k$, there is a diffusive pole in the conductivity but in the limit $k=0$, this pole simply ceases to exist due to a cancellation between factors of $\omega$ in the numerator and denominator and the resultant optical conductivity $\sigma(\omega)$ is a constant.} As $k$ is increased, these poles approach each other until they collide at the crossover point $k\sim\Gamma$ and move off the imaginary axis, forming the two sound-like modes with dispersion relations (\ref{eq:hydrodispersionsofsoundlikemodes}). As $k$ is increased further into the range $\Gamma\ll k\ll\Lambda$, these modes more closely resemble the familiar sound waves (\ref{eq:hydrosounddispersionrelation}) from ordinary hydrodynamics, as we probe shorter and shorter distances at which momentum is better conserved.

We can see the transition from another viewpoint by tracking the poles in the complex frequency plane at a fixed wavenumber $k$, as the momentum dissipation rate $\Gamma$ is increased from zero. When $\Gamma=0$, there are two poles near the real axis, due to the sound excitations with dispersion relations (\ref{eq:hydrosounddispersionrelation}). At small $\Gamma\ll k\ll\Lambda$, these sound-like poles move deeper into the complex frequency plane as their decay rate increases in line with equation (\ref{eq:hydrodispersionsofsoundlikemodes}). As we approach the crossover point $\Gamma\sim k$, each of these poles begins to move towards the imaginary axis until they collide, signifying the crossover, and produce two purely imaginary poles with dispersion relations \eqref{eq:hydrodiffusiondispersion} and \eqref{eq:hydroDrudetypemodefinitek}. The diffusive mode moves towards the origin and becomes more stable, while the other mode moves down the imaginary axis and becomes short-lived. A schematic depiction of this is shown in figure \ref{fig:PicturesQNMS}.

We can identify the crossover point as the point at which the speed of the sound-like modes vanish and they collide. From (\ref{eq:hydrocrossoverpolesanalytic}), this happens when
\begin{equation}
\Gamma k^{-1}+\frac{\eta}{\epsilon+P}k=2\sqrt{\frac{\partial p}{\partial\epsilon}}.
\end{equation}
For a conformal fluid with $\Gamma\ll T$, this implies that $k\sim\Gamma$ at the crossover (due to the scalings described in footnote \ref{footnote3}). This crossover, visible in the poles, is also visible in the full spatially resolved heat conductivity $\kappa\left(\omega,k\right)$.

\subsection{Fast momentum dissipation: the incoherent regime \label{section:incoherent}}

As previously mentioned, when the momentum dissipation rate becomes $\Gamma\sim\Lambda\sim T$, the Drude-like formula (\ref{eq:coherenthydroopticalconductivity}) is no longer accurate and we must consider the influence of other excitations in the system. Because of the microscopic nature of these other excitations, there is no universal theory (such as the hydrodynamics of the previous subsection) that will be applicable when $\Lambda\ll\Gamma$. However, at low frequencies and momenta $\omega,k\ll\Lambda\ll\Gamma$, there are some universal features present.

Again, it is simplest to first consider the AC heat conductivity $\kappa(\omega,k=0)$. For $\Gamma\sim\Lambda$, the potential Drude-like pole has moved deep enough into the complex plane that it has a comparable decay rate to the microscopic collective excitations. This means that no coherent excitations of the system survive at very long timescales $\omega\ll\Gamma\sim\Lambda\sim T$, and the resulting spectrum in this range of frequencies is incoherent. That is, the heat conductivity will have no sharp peaks signifying the existence of a long-lived collective mode. This is clear by examining the structure of the poles of $\kappa(\omega,0)$ in the complex $\omega$ plane: there are no poles within the range $\left|\omega\right|<\Lambda$ and so the AC conductivity $\kappa(\omega,0)$ will be approximately constant over this range of frequencies. This is an example of an incoherent state. With our UV scale $\Lambda_{UV}\sim T$, the conductivity will be approximately constant for all $\omega\lesssim\Lambda_{UV}$. However, if one takes $\Lambda_{UV}\gg\Lambda\sim T$, there will be a large frequency range $\Lambda\ll\omega\ll\Lambda_{UV}$ over which the excitations with decay rates $\sim\Lambda$ will influence the conductivity. The outcome in this case will depend upon the details of the specific theory and its excitations, and is beyond the scope of this paper.

There is also a universal aspect to the spatially resolved heat conductivity $\kappa(\omega,k)$ in the regime of fast momentum dissipation $\Lambda\lesssim\Gamma$. At very long distances $k\ll\Lambda\ll\Gamma$, the microscopic excitations are not important and the resulting physics is governed by conservation laws. As momentum is no longer conserved (not even approximately) when $\Lambda\ll\Gamma$, the only conservation law left is that of energy density, given in equation (\ref{eq:energyconservationfluctuation}). Using $\partial_iP_i=\partial_iQ_i=-\kappa\partial_i\partial_i T$ and elementary thermodynamics, the energy conservation equation may be written as a diffusion equation $\partial_t\varepsilon-D_\parallel\partial_i\partial_i\varepsilon=0,$ with diffusion constant $D_\parallel=\kappa/(T\partial s/\partial T)$, so that the two-point retarded Green's function of the energy density is
\begin{equation}
G^R_{\varepsilon\varepsilon}(\omega,k)=T^2\frac{\partial s}{\partial T}\frac{D_\parallel k^2}{i\omega-D_\parallel k^2}.
\end{equation}
At non-zero wavenumbers $k$, the Ward identity due to energy conservation therefore implies that the thermal conductivity (up to contact terms) takes the form
\begin{equation}
\label{eq:incoherenthydroopticalconductivity}
\kappa(\omega,k)=T\frac{\partial s}{\partial T}D_\parallel\frac{i\omega}{i\omega-D_\parallel k^2}.
\end{equation}
Thus, at the longest distances, heat diffuses in this incoherent state. Furthermore, the DC heat conductivity is controlled by the energy diffusion constant via the Einstein relation
\begin{equation}
\label{eq:incoherenthydroDCconductivity}
\kappa_{DC}=T\frac{\partial s}{\partial T}D_\parallel,
\end{equation} 
and the AC heat conductivity $\kappa(\omega,0)$ is equal to this constant value in this approximation, as anticipated in the preceding paragraph.

These long distance transport properties are very similar to the long distance transport properties in the coherent phase (where momentum dissipates slowly) which were described in the previous section. This is because, as previously described, no matter how slowly momentum dissipates with respect to the decay rate of the other excitations intrinsic to the systems, at very long distances it dissipates significantly and thus energy diffusion is the relevant physics at these distance scales. The difference is that the value of $\kappa_{DC}$ in the coherent phase is related to the momentum dissipation rate (i.e.~to the location of a purely imaginary pole of the AC conductivity in the complex frequency plane) via equations (\ref{eq:coherenthydroopticalconductivity}) and (\ref{eq:coherenthydrodcheatconductivity}), whereas in an incoherent phase it is not. In both cases, $\kappa_{DC}$ is related to the diffusion constant $D_\parallel$ by the same Einstein relation (\ref{eq:incoherenthydroDCconductivity}). 

At shorter distances, the physics of the incoherent state (where momentum dissipates quickly $\Gamma\gg\Lambda$) is totally different to that of the coherent state (where it dissipates slowly $\Gamma\ll\Lambda$). In the coherent state, there is a parametrically large region of wavenumbers $\Gamma\ll k\ll\Lambda$ where momentum is approximately conserved and the resulting collective excitations are sound-like excitations (\ref{eq:hydrodispersionsofsoundlikemodes}) -- the resolutions of the Drude-like pole at non-zero $k$. In the incoherent state, there is no corresponding region of wavenumbers. Instead, when one increases $k$ to $\Lambda\ll k\ll\Gamma$ the energy diffusion rate becomes so large that it is comparable to the decay rate of the microscopic excitations of the system, and there is no universal excitation that controls the properties of the system any more. The transport properties will now depend on the specific microscopic details of the system. Of course, if one takes $\Lambda_{UV}\sim\Lambda\sim T$ as the maximum energy scale one is interested in, these details will not be important and one will find diffusive physics at all wavenumbers $k\lesssim\Lambda_{UV}$.

A brief summary of the transport properties at long distances and low energies in both the coherent and incoherent phases is given in table \ref{fig:summarytransporttable}. At zero $k$, the key difference is that the coherent phase will display a Drude-like peak in the AC conductivity due to the long lifetime of momentum whereas the incoherent phase will not. At non-zero $k$, the key difference is that, at sufficiently high wavenumbers (short distances), the coherent phase transports heat via sound-like waves since momentum is approximately conserved over these distances scales, whereas in the incoherent phase this does not happen.

\begin{table}
\begin{tabular}{| l | l | l |}
\hline
\text{} &\;\;\;Coherent phase: $\Gamma\ll\Lambda$\;\;\;&\;\;\;Incoherent phase: $\Lambda\ll\Gamma$\;\;\;\\
\hline
\;\;\;$k=0$\;\;\;&\;\;\;Drude-like peak in $\kappa(\omega,0)$ \;\;\;&\;\;\;constant $\kappa(\omega,0)$\;\;\;\\
\hline
\;\;\;Small $k$\;\;\;&\;\;\;$k\ll\Gamma\ll\Lambda$: diffusion \;\;\;&\;\;\;$k\ll\Lambda\ll\Gamma:$\;\;\;\\
\cline{2-2}
\;\;\;($k\ll\Lambda$)\;\;\;&\;\;\;$\Gamma\ll k\ll\Lambda$: sound\;\;\;&\;\;\;diffusion\\
\hline
\end{tabular}
\caption{Summary of the general features of the AC and spatially resolved conductivities in the coherent and incoherent regimes of the hydrodynamic model.}
\label{fig:summarytransporttable}
\end{table}

\subsection{Transverse transport}
\label{sec:hydrotransversesection}
We have concentrated thus far on the longitudinal response of the theory: the transport of heat (or momentum density) parallel to the direction of a non-uniform source such as a temperature gradient. Let us now briefly address the transverse response of the theory i.e.~when a non-uniform source is present, how is momentum transported in the direction perpendicular to this non-uniformity? By analogy with the heat conductivity (\ref{eq:hydroheatconductivity}), it is helpful to define the quantity
\begin{align}
\label{eq:transversemomentumconductivitydefninhydro}
\kappa_\perp(\omega,k)\equiv\frac{i}{\omega T}\left[G^R_{P^yP^y}(\omega,k)-G^R_{P^yP^y}(0,k)\right]=\frac{s}{-i\omega+\Gamma+k^2\frac{\eta}{\epsilon+p}},
\end{align}
which we will call the transverse momentum conductivity. In the limit $k=0$ in which rotational invariance is restored, this coincides with the usual AC heat conductivity. To determine the final equality in equation (\ref{eq:transversemomentumconductivitydefninhydro}), we have used the results (\ref{eq:transverseGreensfunctioninhydro}) for the retarded Green's functions of a hydrodynamic theory with slow momentum dissipation.

It is worthwhile to investigate the properties of $\kappa_\perp(\omega,k)$ in both the incoherent and coherent phases. Due to rotational invariance, at $k=0$ the properties are identical to those of the heat conductivity: there is a purely imaginary pole in the coherent phase ($\Gamma\ll\Lambda$) which produces a Drude-like peak in the AC conductivity, and no long-lived poles in the incoherent phase ($\Lambda\ll\Gamma$) leading to an approximately constant AC conductivity at the lowest frequencies $\omega\ll\Lambda$.

At non-zero wavenumbers, $\kappa_\perp$ differs from the heat conductivity. Consider first the case where there is no momentum dissipation: $\Gamma=0$. This is just ordinary hydrodynamics, in which the conservation of momentum ensures that transverse momentum density diffuses with a diffusion constant $D_\perp=\eta/\left(\epsilon+p\right)$. This can be seen clearly in the poles of $\kappa_\perp(\omega,k)$ at $\Gamma=0$: the zero frequency pole at $k=0$ -- which produces the delta function in the heat conductivity -- is resolved into a diffusive mode in the transverse momentum conductivity at non-zero $k$. 

In the coherent case ($\Gamma\ll\Lambda$), as the wavenumber $k$ is increased from zero, the Drude-like excitation begins to dissipate faster, and the pole moves deeper into the complex frequency plane with dispersion relation
\begin{equation}
\label{eq:transversemomentumdisprelhydro}
\omega=-i\Gamma-i\frac{\eta}{\epsilon+p}k^2+\ldots.
\end{equation}
When $k$ becomes large enough, the lifetime of this pole will become similar to $\Lambda$, the lifetime of the microscopic excitations, at which point this hydrodynamic analysis is no longer valid and one must deal directly with the specific details of individual theories. 

In the incoherent case ($\Lambda\lesssim\Gamma$), there are no long-lived poles near the origin at $k=0$ as transverse momentum is no longer approximately conserved. At distances $k\ll\Lambda$, the transverse momentum conductivity will be featureless, and at shorter distances the microscopic details of the system will be important. Finally, let us emphasise here that the diffusion of transverse momentum when $\Gamma=0$ is totally unrelated to the diffusion of longitudinal momentum when $\Gamma\ne0$: the former arises due to momentum conservation while the latter is due to energy conservation, and both have different diffusion constants.

\section{Heat transport in a neutral axionic theory \label{section:AxionTheory}}

\subsection{The holographic model}

We now focus our attention on a specific, strongly interacting thermal state, and examine how it transports heat and momentum. We will find that it exhibits both a coherent and an incoherent phase, and that the effective description just outlined is a very good model for its low energy and long distance properties. This (2+1)-dimensional, strongly interacting state is holographically dual to a solution of the action
\begin{equation}
\label{eq:axiontheoryaction}
S=\int d^4x \sqrt{-g}\left(\mathcal{R}+6-\frac{1}{2}\sum_{i=1}^{2}\partial^\mu\phi_i\partial_\mu\phi_i\right),
\end{equation}
of gravity, with a negative cosmological constant, coupled to two massless scalar fields. When $\phi_i=0$, this theory has a planar Schwarzschild-AdS$_4$ black brane solution, which is dual to a field theory state that behaves hydrodynamically due to the conservation of energy and momentum. To obtain non-trivial results for heat transport, we must incorporate a mechanism by which momentum can dissipate. This is precisely the role of the neutral scalar fields: this action has a very simple planar black brane solution in which the scalar fields explicitly break translational invariance and thus momentum, in the dual field theory, is no longer conserved \cite{Bardoux:2012aw,Andrade:2013gsa}
\begin{align}
\label{eq:axiontheorymetricsolution}
ds^2&=-r^2f(r)dt^2+r^2\left(dx^2+dy^2\right)+\frac{dr^2}{r^2f(r)},\;\;\;\;\;\;\phi_1=mx,\;\;\;\;\;\;\phi_2=my,\\
f(r)&=1-\frac{m^2}{2r^2}-\frac{r_0^3}{r^3}\left(1-\frac{m^2}{2r_0^2}\right). \nonumber
\end{align} 
The radial co-ordinate $r$ can take values between $r=r_0$ (the location of the black brane horizon) and $r=\infty$ (the asymptotically AdS boundary), and the co-ordinates $t,x,y$ label the directions of the dual field theory. The scalar fields break translational symmetry in such a simple way that the resulting spacetime is isotropic and homogeneous. 

The thermodynamic properties of this field theory state were investigated in \cite{Bardoux:2012aw,Andrade:2013gsa} and we summarise them here for future reference. The temperature $T$, energy density $\epsilon$, pressure $p$ and entropy density $s$ of the state are
\begin{equation}
\label{eq:thermodynamicsofaxiontheory}
T=\frac{r_0}{4\pi}\left(3-\frac{m^2}{2r_0^2}\right),\;\;\;\;\;\;\epsilon=2r_0^3\left(1-\frac{m^2}{2r_0^2}\right),\;\;\;\;\;\;p=r_0^3\left(1+\frac{m^2}{2r_0^2}\right),\;\;\;\;\;\;s=4\pi r_0^2.\;\;\;\;\;\;
\end{equation}
We restrict the parameter range to $0\le m\le\sqrt{6}\,r_0$ so that the temperature is never negative. At zero temperature, the IR geometry becomes $AdS_2\times R^2$, similar to the charged AdS Reissner-Nordstr\"om black brane with translation invariance. Observe though that at low $m$, the energy density is positive, and decreases as $m$ increases, eventually becoming negative. The value of $m=\sqrt2\,r_0$ where it vanishes will be of special interest as we will shortly see. At this value, all the components of the holographic energy-momentum tensor vanish, $\langle T_{\alpha\beta}\rangle=0$, where $\alpha,\beta=t,x,y$, and the metric is conformal to a patch of AdS$_2\times R^2$. This also bears some similarity to hyperbolic black holes (note that the $m^2$ term in $f(r)$ acts like a negative horizon curvature), \cite{Emparan:1999gf}, which also have a zero energy density state with nonzero entropy and temperature. However, it turns out to be isometric to AdS space, and the horizon is analogous to a Rindler acceleration horizon. In our case though, the zero energy density state is still described by a black hole, with a bifurcate Killing horizon and a curvature singularity.

The solution (\ref{eq:axiontheorymetricsolution}) satisfies the null energy condition for all $m$. The action \eqref{eq:axiontheoryaction} can easily be generalised to allow for IR geometries with other values of the dynamical exponent $z$ and hyperscaling violating exponent $\theta$ \cite{Charmousis:2010zz,Gouteraux:2011ce,Gouteraux:2014hca}. 

As usual, the metric $g_{\mu\nu}$ of the gravitational theory is dual to the energy-momentum tensor of the dual field theory. The extra `axionic' scalar fields (their background values are odd under a parity transformation $(x,y)\rightarrow(-x,-y)$) are dual to neutral scalar operators in the dual field theory, which have spatially dependent sources that break translational invariance. The coefficient $m$ is the UV parameter controlling the breaking of translational invariance and is connected to the rate of momentum dissipation in the dual field theory. When studying linearised excitations around this black brane solution, $m$ acts as an effective mass for the graviton, and there is a very close connection between `axionic' black brane solutions and black brane solutions of theories of massive gravity \cite{Andrade:2013gsa,Taylor:2014tka}.

In the following subsections, we will compute the conductivities $\kappa(\omega,k)$ and $\kappa_\perp(\omega,k)$ of the state for different values of $m$ (which will control whether the state behaves in a coherent or incoherent way). These conductivities are computed from the retarded Green's functions, which we principally determine numerically from a linearised analysis of perturbations around the black brane solution. The principles behind such calculations are fairly standard, and we refer the interested reader to appendix \ref{sec:numericalappendix} for a description of the technical details of our calculations. In the following, we will also take the UV scale to be $\Lambda_{UV}\sim T$ i.e.~we will investigate the linear response in the range $\omega,k \ll T$. We note that there could be interesting and non-trivial physics within the region $T\ll\omega,k\ll\Gamma$, but we leave this to be addressed in the future.

\subsection{The AC heat conductivity \texorpdfstring{$\kappa(\omega)$}{}}

We begin our analysis of the transport properties with the AC heat conductivity at zero wavenumber $\kappa(\omega)\equiv\kappa(\omega,k=0)$, which controls the heat transport at the very longest distance scales. The DC limit of the heat conductivity, $\kappa_{DC}\equiv\kappa(\omega\rightarrow0)$, can be calculated analytically \cite{Donos:2014cya} and is found to be 
\begin{equation}
\label{eq:analytickappadcequation}
\kappa_{DC}=\frac{4\pi sT}{m^2},
\end{equation}
for any value of $m$.\footnote{Although this is very reminiscent of the corresponding memory matrix expression \cite{forster1990hydrodynamic, Hartnoll:2007ih, Hartnoll:2012rj, Mahajan:2013cja}, it differs at large $m/T$ where the entropy density in the numerator of \eqref{eq:analytickappadcequation} is significantly different from $s\left(m=0\right)$.} For \textit{sufficiently small values of $m$}, previous analyses have shown that this system behaves coherently with a momentum dissipation rate \cite{Davison:2013jba,Andrade:2013gsa}
\begin{equation}
\Gamma=\frac{s}{4\pi\left(\epsilon+p\right)}m^2=\frac{m^2}{4\pi T}.
\label{eq:analyticgammaholo}
\end{equation} 
That is, for small enough $m$ there is a Drude-like pole in the AC conductivity at $\omega=-im^2/4\pi T$ that controls the low energy heat transport in this system. The DC conductivity (\ref{eq:analytickappadcequation}) is consistent with the result (\ref{eq:coherenthydrodcheatconductivity}) of our coherent hydrodynamic model with the dissipation rate (\ref{eq:analyticgammaholo}). However, the analytic result \eqref{eq:analytickappadcequation} for $\kappa_{DC}$, true for \textit{any} $m$, makes it tempting to speculate that a hydrodynamic description with $\Gamma=m^2/4\pi T$ is always valid and, therefore, that this is a coherent state for any $m$. This is not the case, as we will explicitly show. The reason for this can be understood on general grounds: the hydrodynamic theory has an intrinsic energy scale $\Lambda\sim T$ characterising the decay rates of its microscopic excitations. Since momentum has a decay rate $\Gamma\sim m^2/T$ at small $m$, it is clear that when $m$ is increased enough such that $m\sim T$, the decay rate of momentum is of the same order as that of the microscopic collective excitations $\Gamma\sim\Lambda$ and we therefore pass from a hydrodynamic, coherent phase, where the low energy physics is controlled solely by momentum conservation, to an incoherent phase where there is no single long-lived quantity in control. This serves as a warning that it is not possible to infer the characteristics of AC transport (such as whether there is a Drude peak or not) from DC quantities alone.
\begin{figure}
\begin{tabular}{cc}
\includegraphics[width=.45\textwidth]{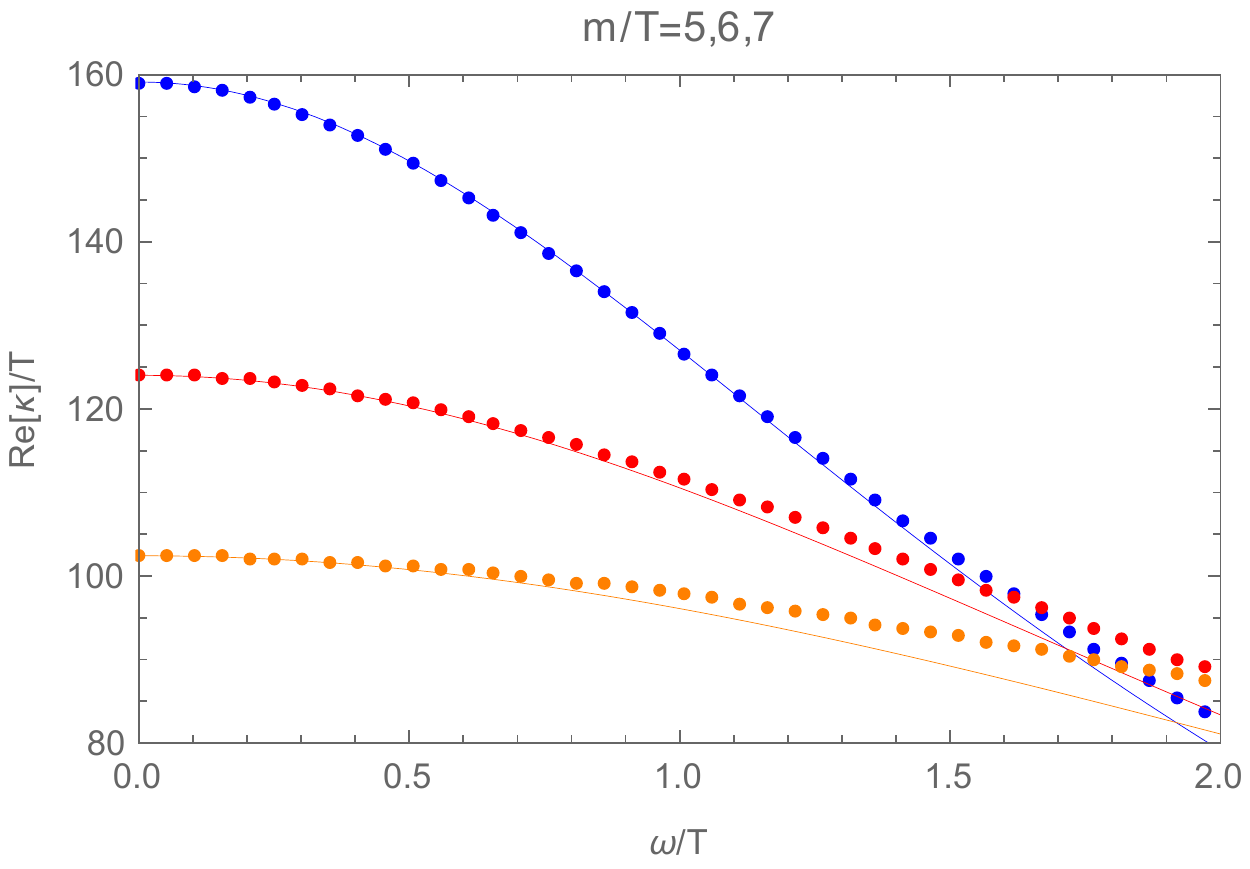}&
\includegraphics[width=.45\textwidth]{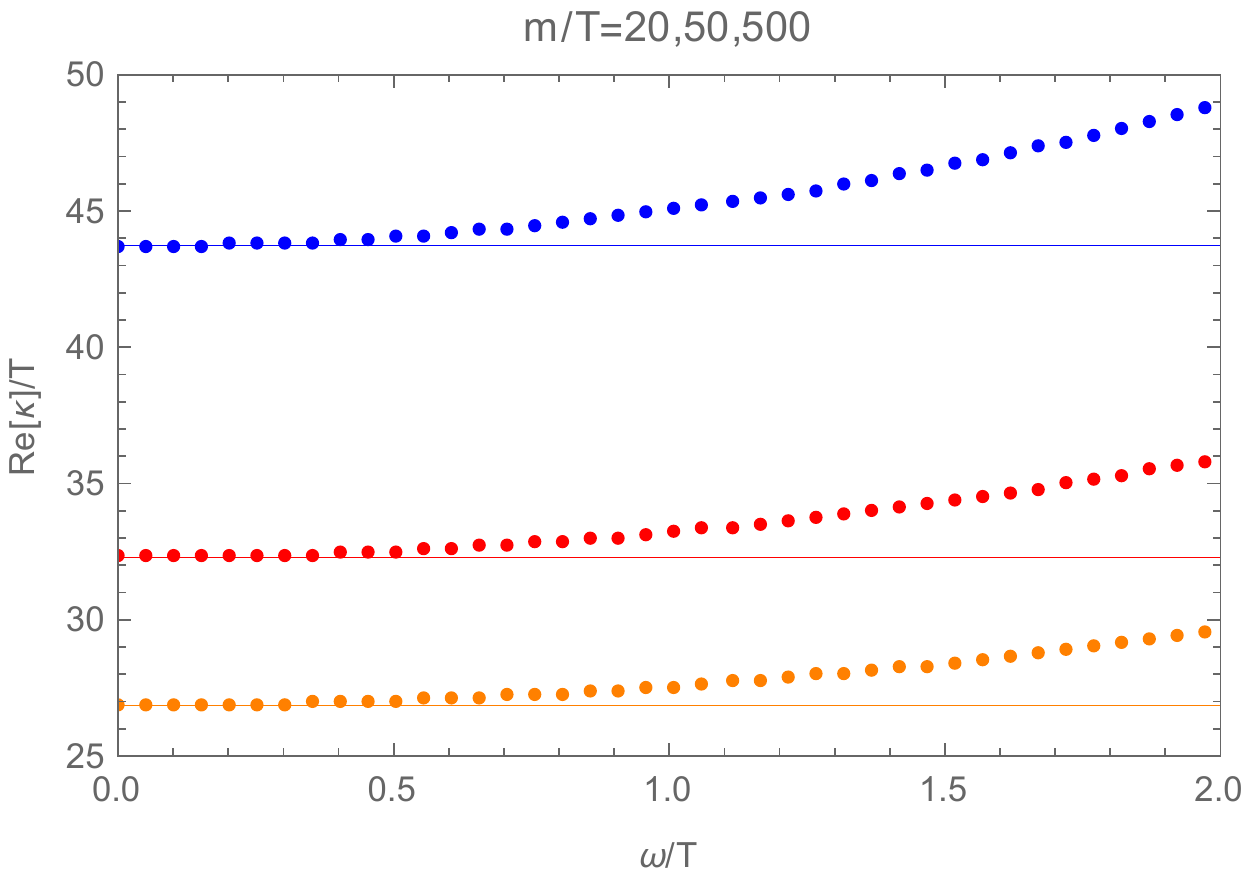}
\end{tabular}
\caption{Plots of the AC heat conductivity at low (left panel) and high (right panel) values of $m/T$. Values of $m/T$ (given in the plot headings) increase from top to bottom. The dots are the numerical results of our holographic model, and the solid lines are the analytical results \protect\eqref{eq:coherenthydroopticalconductivity} (left panel) and \protect\eqref{eq:incoherenthydroopticalconductivity} (right panel) of the hydrodynamic model. The agreement with the Drude-like conductivity \protect\eqref{eq:coherenthydroopticalconductivity} is better at lower values of $m/T$ (coherent regime, left), while the constant conductivity \protect\eqref{eq:incoherenthydroopticalconductivity} is better at higher values of $m/T$ (incoherent regime, right). Note also the change of slope from peak (left) to valley (right).}
\label{fig:ConductivityNeutralk=0}
\end{figure}

Upon examining the low frequency AC heat conductivity of the holographic model (shown in figure \ref{fig:ConductivityNeutralk=0}), we find an excellent agreement with the Drude-like form (\ref{eq:coherenthydroopticalconductivity}) associated with coherent transport, with $\Gamma$ given by (\ref{eq:analyticgammaholo}), for $m/T\ll1$. Near $m/T\sim1$, the Drude-like peak disappears, indicating a crossover from a coherent to an incoherent regime. For $m/T\gg1$, a Drude-like peak is not a good fit as there is a valley in the conductivity: at low frequencies, there is a good agreement with the constant conductivity of the hydrodynamic model (\ref{eq:incoherenthydroopticalconductivity}), indicating that neither of these models has  any collective excitations with decay rates $\Lambda\lesssim T$ when $k=0$. At higher frequencies, the models begin to differ due to the existence of higher energy excitations in the holographic system, which are not accounted for in the hydrodynamic effective theory. Note that the coherent/incoherent crossover is gradual, and it is not possible to isolate a single value of $m/T$ at which it occurs. As an order of magnitude estimate, in our holographic model it occurs around $m/T\sim10$.

\begin{figure}
\begin{tabular}{cc}
\includegraphics[width=.45\textwidth]{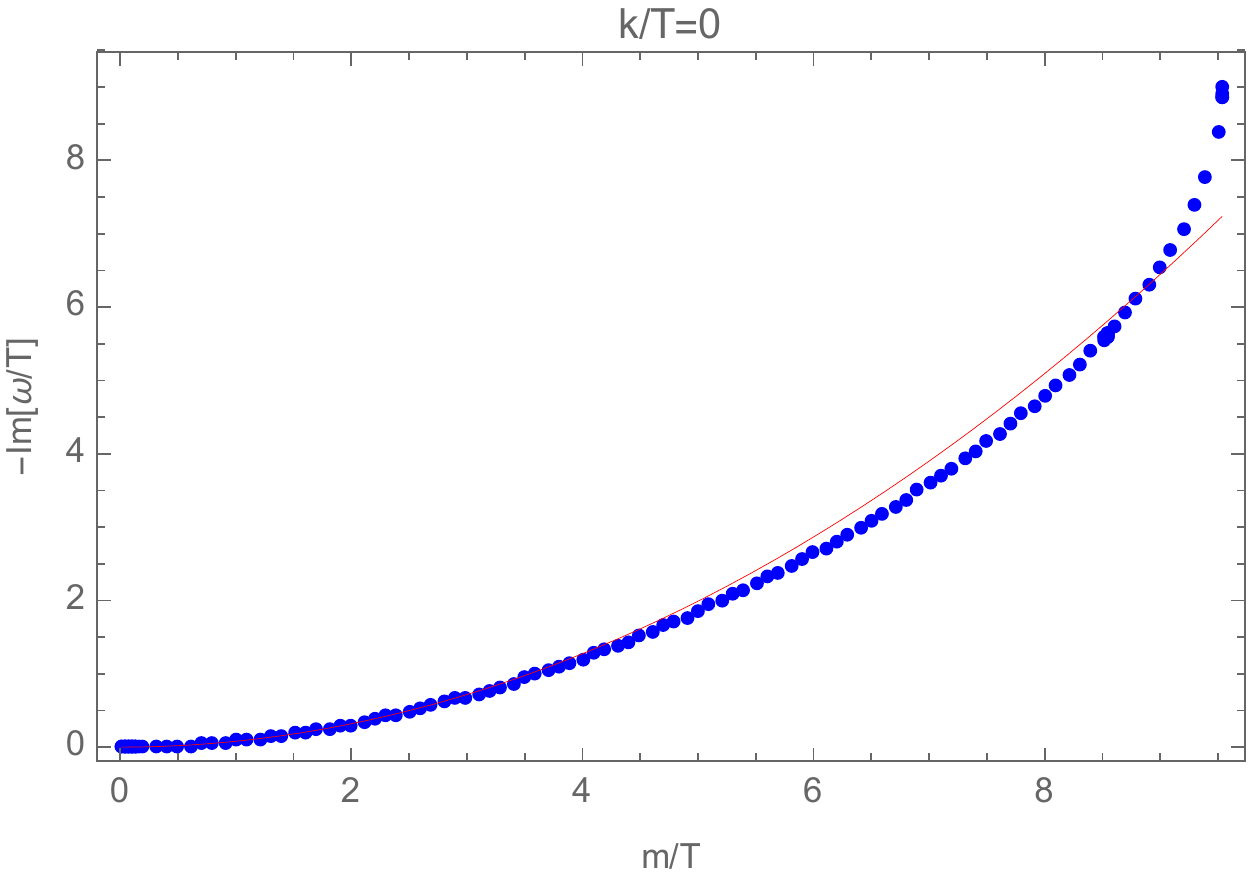}&
\includegraphics[width=.45\textwidth]{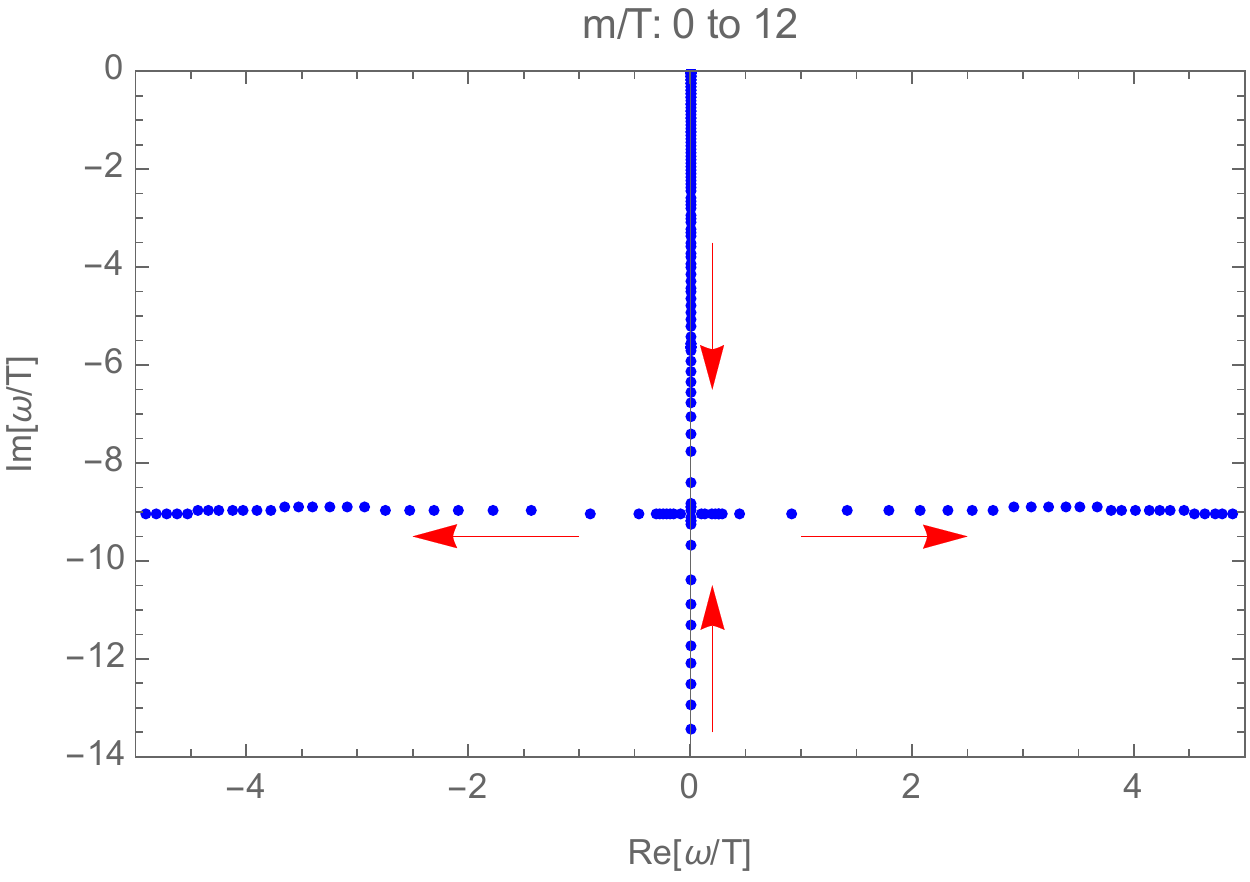}
\end{tabular}
\caption{Poles of the AC conductivity of the holographic model, as $m/T$ is increased from the coherent to incoherent regimes. The left panel shows the purely imaginary pole closest to the origin in our holographic model (dots), compared to the prediction $\omega=-im^2/4\pi T$ (solid line) for a state whose momentum slowly dissipates at a constant rate $m^2/4\pi T$. There is excellent agreement when $m\ll T$, indicating that this is a coherent state with the aforementioned dissipation rate, but there is disagreement for $m\gtrsim T$ as the crossover to an incoherent regime begins. When $m\sim9.5T$, the pole is no longer purely imaginary and we have definitively exited the coherent regime. The right panel shows the location of the poles of the holographic model in the complex frequency plane. The arrows show the movements of the poles as $m$ is increased. At small $m$, there is only a Drude-like pole near the origin. As $m$ increases, this moves down the imaginary axis before colliding with another purely imaginary pole when $m\sim9.5T$, producing two off-axis (non-Drude-like) poles.}
\label{fig:zerokpoles}
\end{figure}

A complementary perspective on these results is found by examining the analytic structure of the heat conductivity $\kappa\left(\omega\right)$ in the complex frequency plane, shown in figure \ref{fig:zerokpoles}. 
From equation (\ref{eq:coherenthydroopticalconductivity}), we know that in the translationally invariant case ($m=0$), the heat conductivity has a pole at zero frequency. As $m/T$ is increased, this pole moves down the imaginary $\omega$ axis. For small values of $m\ll T$, there is no other pole near the origin of the complex $\omega$ plane (i.e. with $|\omega|\ll\Lambda\sim T$), and the low frequency conductivity is described very well by the Drude-like result (\ref{eq:coherenthydroopticalconductivity}) with $\Gamma$ given by (\ref{eq:analyticgammaholo}). For small $m/T$, the relevant effective theory is just the hydrodynamic model of section \ref{section:HeatTransportGen}, and the location of this pole at $\omega=-i\Gamma$ implies a momentum dissipation rate $\Gamma=m^2/4\pi T$. As $m$ is increased to $m\sim T$, this pole moves further from the origin such that its decay rate becomes of the same order of magnitude as that of the other collective excitations of the theory ($\Gamma\sim T$), and the low energy physics is therefore not controlled by slow momentum dissipation: we have entered an incoherent regime. As these poles all have imaginary parts $\sim T\sim\Lambda_{UV}$, there are no poles near the origin of the complex $\omega$ plane, and the conductivity is therefore approximately constant at small $\omega\ll T$. At even larger $m\gg T$, the short-lived pole which was originally Drude-like combines with another imaginary pole and forms two short-lived, propagating excitations. This makes it abundantly clear that a Drude-like response is not applicable here and that no almost-conserved quantity exists. Furthermore, there is no way to extract a momentum dissipation rate $\Gamma$ from the AC conductivity in this incoherent region, as there is no Drude-like pole.

We stress this point: despite the fact that the DC conductivity (\ref{eq:analytickappadcequation}) is consistent with coherent (i.e.~Drude-like) transport with momentum dissipation rate (\ref{eq:analyticgammaholo}) for all $m/T$, there is actually incoherent transport when $m/T\gg1$. The fact that a naive extrapolation of the Drude-like result for the DC conductivity in the coherent regime also gives the correct DC conductivity in the incoherent regime is likely a result of the simplicity of our holographic theory, and should not be true in general. This is apparent from the hydrodynamic analysis in section \ref{section:HeatTransportGen}: in the coherent regime, the DC conductivity is determined by the momentum dissipation rate \eqref{eq:coherenthydrodcheatconductivity}, while in the incoherent regime it is determined by an independent object -- the energy diffusion constant \eqref{eq:incoherenthydroDCconductivity}.

Finally, we note that in this holographic system there is a special value of the translational symmetry breaking parameter $m=\sqrt{8}\pi T$ for which $\kappa\left(\omega\right)$ is exactly constant for all $\omega$, due to the coincidence of its poles and zeroes in the complex $\omega$ plane. This is not generic, and is a special feature of our holographic model associated with self-duality of the gravitational theory. At this value, the energy density black brane vanishes. We postpone a more detailed discussion of this to section \ref{section:SelfDuality}.

\subsection{Spatially resolved transport}

Having established the existence of a coherent regime (when $m\ll T$) and an incoherent regime (when $m\gtrsim T$) from the spatially uniform response of the system, we will now examine the spatially resolved conductivities, and determine the characteristic low energy excitations that transport heat in each regime.

\subsubsection{Longitudinal response\texorpdfstring{: $\kappa(\omega,k)$}{}}

The longitudinal response -- the response of the momentum (or equivalently heat current) to a source spatially modulated in the direction of momentum flow -- is controlled by the usual heat conductivity $\kappa\left(\omega,k\right)$, given by the Kubo formula in equation (\ref{eq:hydroheatconductivity}). We again find qualitatively different behaviours in the coherent and incoherent regimes, consistent with the analysis in section \ref{section:HeatTransportGen}. 

\begin{figure}
\begin{tabular}{cc}
\includegraphics[width=.45\textwidth]{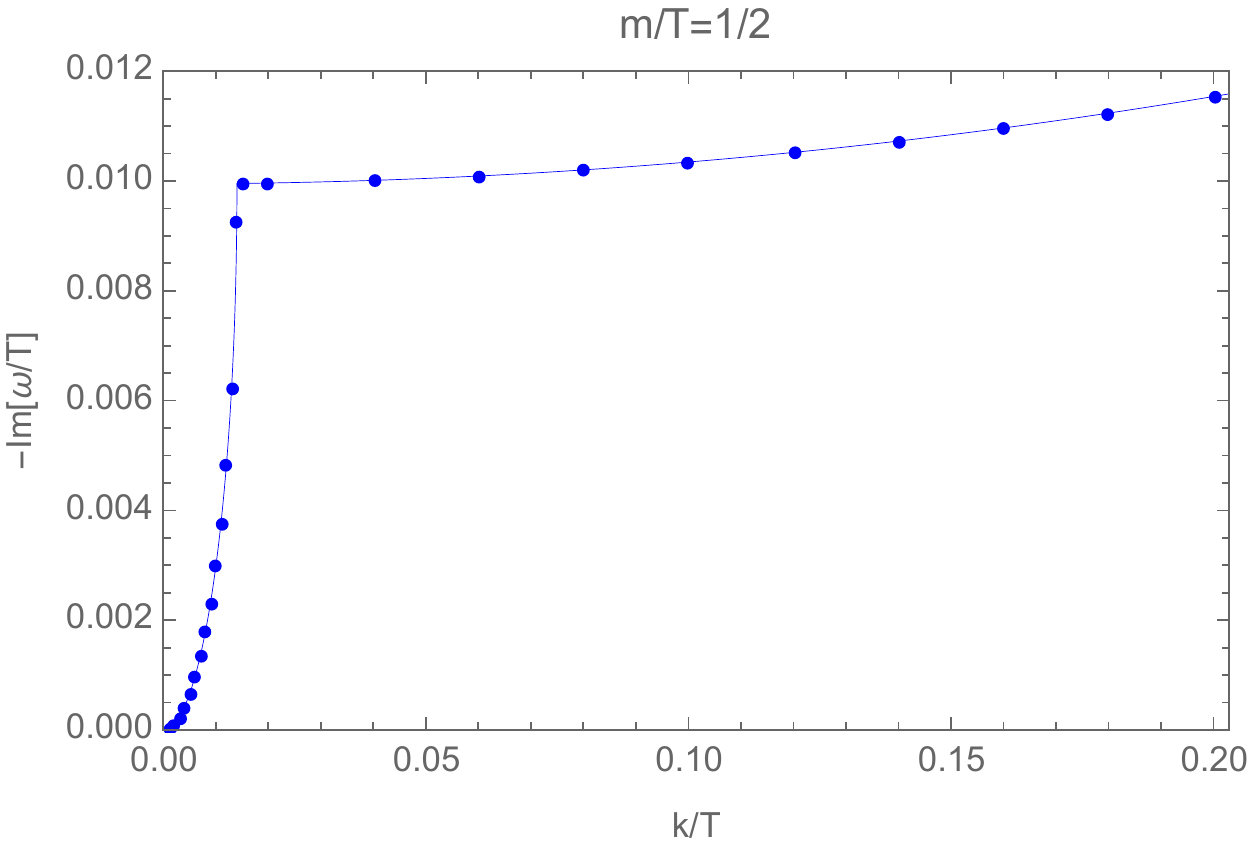}&
\includegraphics[width=.45\textwidth]{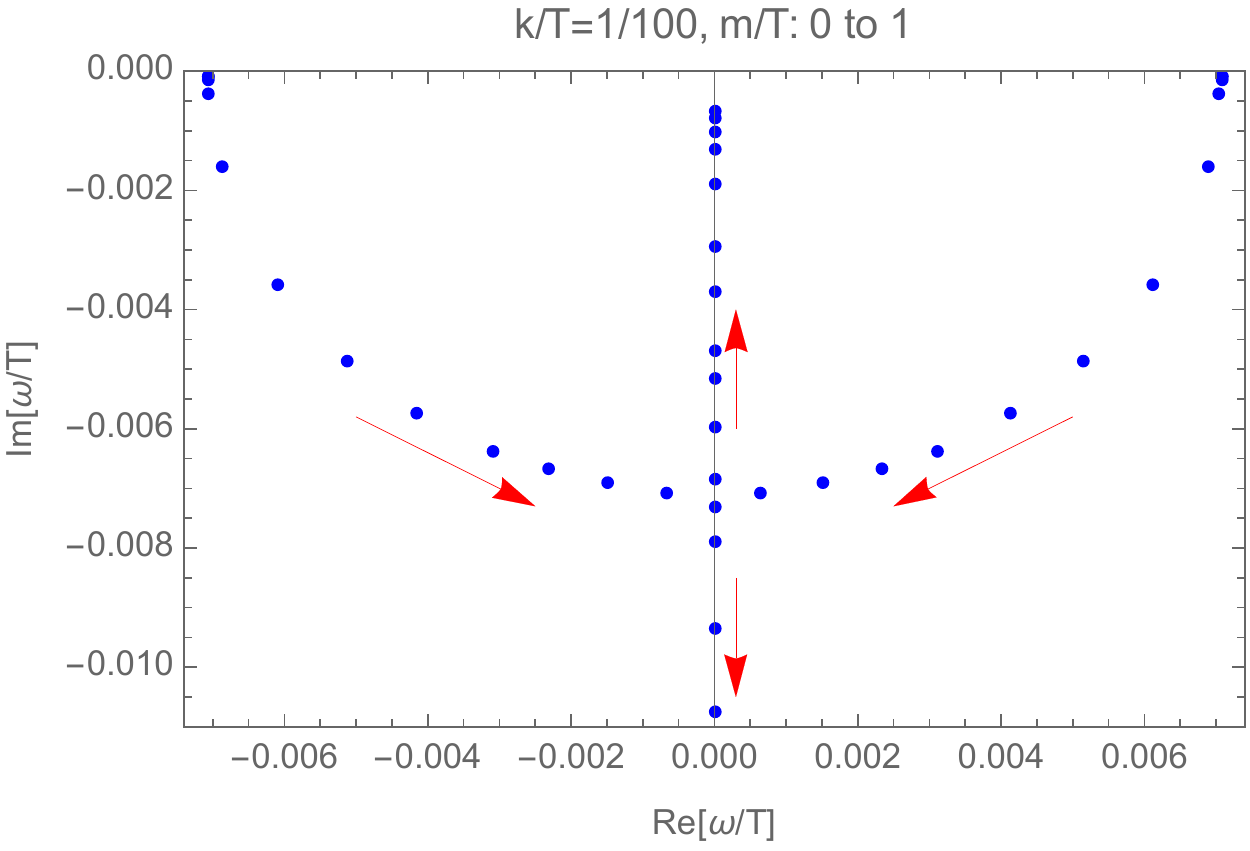}
\end{tabular}
\caption{The crossover between diffusive transport ($k\ll\Gamma\ll T$) and sound-like transport ($\Gamma\ll k\ll T$) in the coherent regime. The left panel shows the imaginary part of the dispersion relation of the longest lived excitation of the holographic model (dots), and the corresponding dispersion relation (\protect\ref{eq:coherentcombinedexcitationholo}) in the hydrodynamic model (solid line), when $m/T=1/2$. There is excellent agreement, indicating diffusion (\protect\ref{eq:holodiffusionanalytic}) at small $k$ and a sound-like excitation (\protect\ref{eq:holosoundlikeanalytic}) at large $k$. The right panel shows the movement of the long-lived poles of the holographic model in the complex $\omega$ plane, as $m$ is increased (within the coherent regime) at fixed $k/T=1/100$. The arrows show the motion of the poles as $m$ increases: at small $m$ (i.e. $\Gamma\ll k$), there are sound-like excitations near the real axis. As $m$ increases, these poles approach each other and then collide at $\Gamma\sim k$, forming a diffusive excitation that moves up the imaginary axis and is long-lived when $k\ll\Gamma$.}
\label{fig:ComplexWPlanePolesHolo}
\end{figure}

In the coherent regime ($m\ll T$), the spatially uniform ($k=0$) conductivity has one dominant pole near the origin of the complex frequency plane (i.e. one long-lived collective excitation): the Drude-like pole at $\omega=-i\Gamma$. According to the hydrodynamic analysis in section \ref{section:HeatTransportGen}, at non-zero $k$ this is resolved into a pole corresponding to an excitation with dispersion relation (\ref{eq:hydrocrossoverpolesanalytic}), whose character depends upon the distance scale under consideration. At small $k\ll\Gamma\ll T$, the longest-lived excitation should be diffusive, with dispersion relation (\ref{eq:hydrodiffusiondispersion}), while the longest-lived excitations at large $k$ ($\Gamma\ll k\ll T$) should be sound-like waves with dispersion relation (\ref{eq:hydrodispersionsofsoundlikemodes}). Using the thermodynamic formulae (\ref{eq:thermodynamicsofaxiontheory}) for our state, along with the expression \eqref{eq:analyticgammaholo} for the momentum dissipation rate in the coherent regime, the dispersion relation (\ref{eq:hydrocrossoverpolesanalytic}) is
\begin{equation}
\label{eq:coherentcombinedexcitationholo}
\omega=\pm\sqrt{\sqrt{\frac{1}{4}+\frac{3m^2}{32\pi^2 T^2}}-\frac{\left(m^2+k^2\right)^2}{64\pi^2 k^2T^2}}\,k-\frac{i}{8\pi T}\left(m^2+k^2\right)+\ldots,
\end{equation}
where we have used $\eta=s/4\pi$ (this is only true when $m=0$, but in the coherent regime the $m^2$ corrections to this quantity are subleading and so may be safely ignored). In the relevant limits of small and large $k$, this dispersion relation is either that of a diffusive or a sound-like excitation respectively
\begin{align}
\omega&=-i D_\parallel\,k^2+O(k^4),\qquad D_\parallel=\frac{T}{m^2}\sqrt{4\pi^2+\frac{3m^2}{2T^2}}\,,\label{eq:holodiffusionanalytic}\\
\omega&=\pm\left(\frac{1}{4}+\frac{3m^2}{32\pi^2T^2}\right)^{\frac{1}{4}}k-\frac{i}{8\pi T}\left(m^2+k^2\right)+O(k^3).\label{eq:holosoundlikeanalytic}
\end{align}
In the left hand panel of figure \ref{fig:ComplexWPlanePolesHolo}, we plot the imaginary part of the pole closest to the origin of the complex frequency plane as a function of $k$, confirming the dispersion relations above. The real part of the sound-like excitation (not shown), when it exists, is also consistent with the dispersion relation (\ref{eq:holosoundlikeanalytic}).
To study the transition between sound-like and diffusive transport in the coherent regime, it is convenient to look at the locations of the near-origin poles in the complex frequency plane as a function of $k$: as $k$ is increased from $0$, the diffusive mode emerges from the origin while the Drude-like mode becomes more stable, and moves up the imaginary axis towards the origin. When $k\sim\Gamma$, these poles collide and move off-axis, turning into the long-lived propagating sound-like modes. All of this is consistent with the hydrodynamic analysis of section \ref{section:HeatTransportGen}. This crossover can also be seen by fixing $k$ and increasing $\Gamma$, as shown in the right hand panel of figure \ref{fig:ComplexWPlanePolesHolo}. At $\Gamma=0$, there is perfect momentum conservation and the long-lived excitations are sound modes. At small $\Gamma\ll k$, these become sound-like modes with the dispersion relation (\ref{eq:holosoundlikeanalytic}), where the extra $k$-independent part of their decay rate arises due to the constant rate of momentum loss in the coherent state. Near $\Gamma\sim k\ll\Lambda$, the sound-like poles move towards the imaginary axis and collide, producing two purely imaginary modes. One of these moves away from the origin while one is diffusive and moves towards the origin, in agreement with the dispersion relation (\ref{eq:holodiffusionanalytic}). As $\Gamma$ is increased further such that $\Gamma\gtrsim\Lambda$ (not shown in the figure), the system enters an incoherent regime.

\begin{figure}
\begin{center}
\includegraphics[width=.45\textwidth]{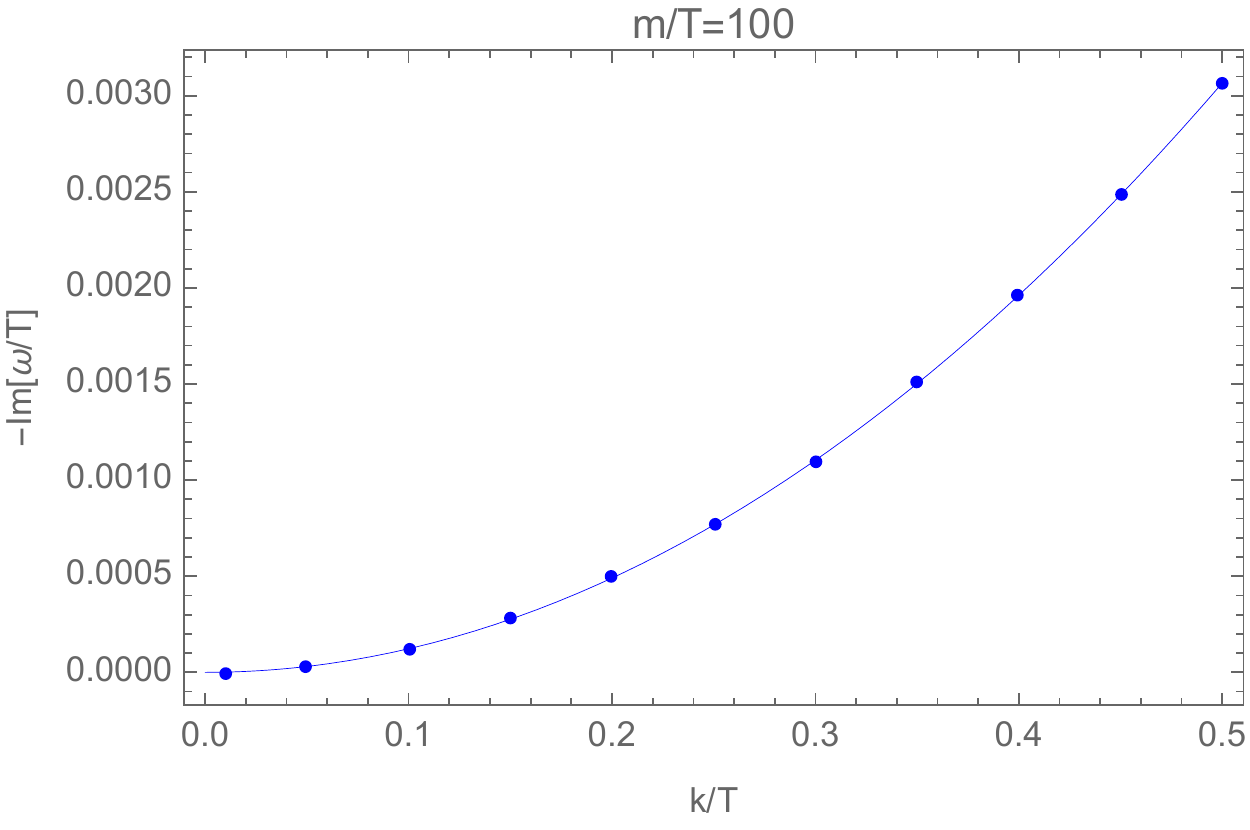}
\end{center}
\caption{The imaginary part of the dispersion relation of the longest-lived pole in the incoherent regime with $m/T=100$ (the real part is zero). The dots denote the poles of the holographic model and the solid line is the analytic result (\ref{eq:holodiffusionanalytic}) of the hydrodynamic model. There is excellent agreement, showing that heat always diffuses in this regime.}
\label{fig:HoloDiffusionIncoherent}
\end{figure}

In the incoherent regime ($m\gg T$), the spatially uniform ($k=0$) conductivity does not have any poles near the origin of the complex frequency plane, as there are no almost-conserved quantities in this system. As described in section \ref{section:HeatTransportGen}, we therefore expect diffusive transport with the dispersion relation (\ref{eq:holodiffusionanalytic})\footnote{Due to the Einstein relation \eqref{eq:incoherenthydroDCconductivity}.} at all distance scales $k\ll T$. This is what we find: there is always only one pole close to the origin for $k\ll T$, and it has the dispersion relation (\ref{eq:holodiffusionanalytic}) as shown in figure \ref{fig:HoloDiffusionIncoherent}. There is no low energy pole collision as there was in the coherent case, because there is no Drude-like pole for the diffusive pole to collide with.

\begin{figure}
\begin{center}
\includegraphics[width=0.5\textwidth]{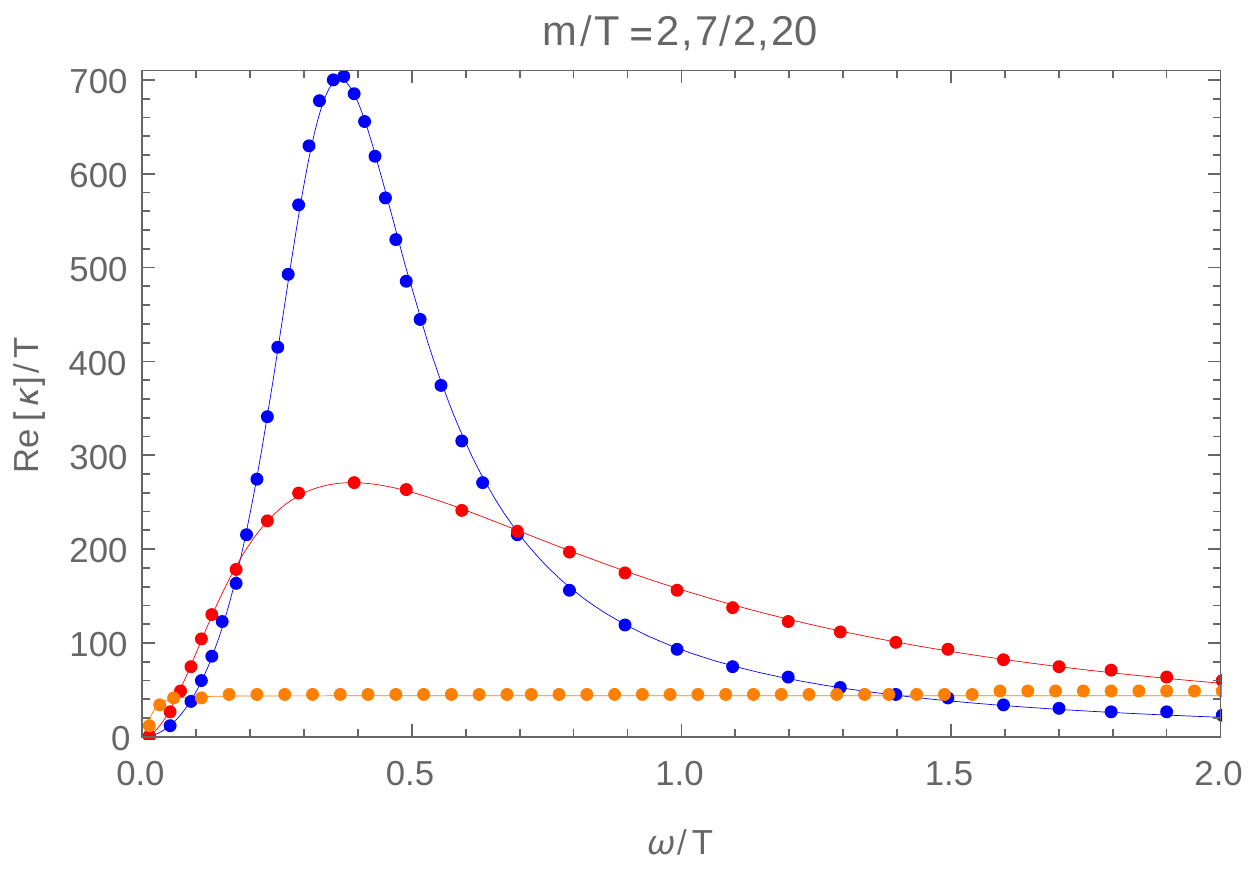}
\end{center}
\caption{Plots of the spatially resolved heat conductivity for $k/T=1/2$, as $m/T$ is increased through the coherent/incoherent transition. Values of $m/T(=2,7/2,20)$ increase from top to bottom. The dots are the numerical results from our holographic model and the solid lines are the predictions of the hydrodynamic model \protect\eqref{eq:hydroheatconductivity} ($m/T=2,7/2$) and \protect\eqref{eq:incoherenthydroopticalconductivity} ($m/T=20$), with $\eta$ approximated as $s/4\pi$ in the coherent regime. There is a crossover from a sound-like excitation at $m/T=2$ to diffusion at $m/T=20$.}
\label{fig:ACHeatConductivityKDependent}
\end{figure}

We have restricted so far to just showing the poles of the Green's functions which correspond to the longest lived excitations of the strongly interacting state. For this to be a sensible approach, we should be able to understand the low energy transport purely in terms of the small number of excitations which comprise our effective theory of transport. In figure \ref{fig:ACHeatConductivityKDependent}, we show the dissipative part of the thermal conductivity $\kappa$ as $\Gamma$ is increased through the coherent/incoherent crossover. The exact (numerical) results are in excellent agreement with the hydrodynamic model of section \ref{section:HeatTransportGen}, showing that this effective theory, that incorporates only the longest-lived excitations, is sufficient to understand the transport properties of the state at long times and distances
There is a clear crossover visible in the plots: in the coherent regime ($m/T=2$), the conductivity is dominated by a sharp peak due to the existence of the sound-like excitations. As $m/T$ is increased through the transition, this peak is washed out as diffusion takes over as the dominant mechanism of charge transport at this distance scale. At higher values of $\omega$ and $k$ (not shown), deviations from the hydrodynamic model occur due to the influence of higher energy excitations that are not incorporated in this effective theory.

\subsubsection{Transverse response\texorpdfstring{: $\kappa_\perp(\omega,k)$}{}}

\begin{figure}
\begin{center}
\includegraphics[width=0.5\textwidth]{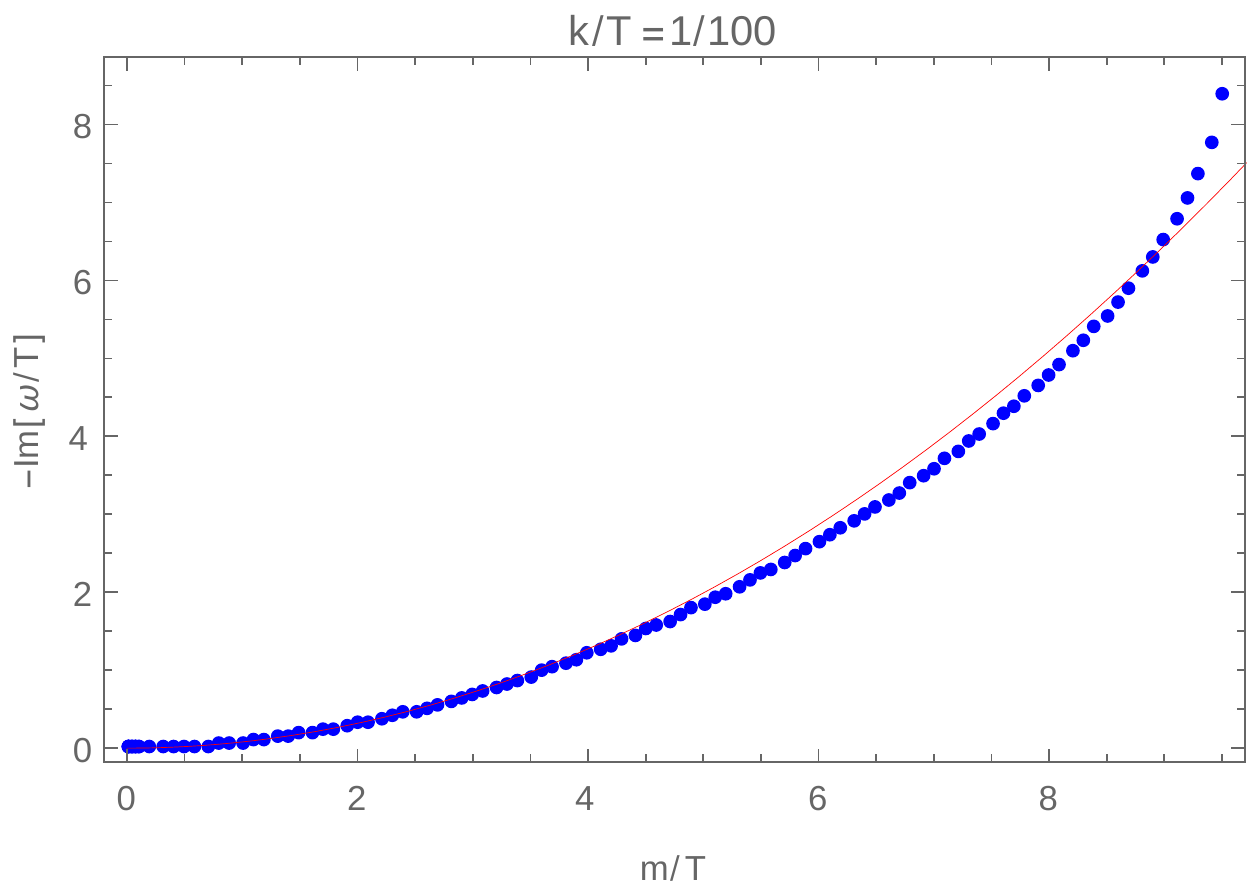}
\end{center}
\caption{The pseudo-diffusive pole of the transverse momentum conductivity as $m/T$ is increased at fixed $k/T=1/100$. The dots are the numerical results for our holographic model, and the solid line is the prediction (\ref{eq:transversemomentumdisprelhydro}) of the hydrodynamic model. For $m\ll T$ there is excellent agreement. For $m\gtrsim T$, as the coherent/incoherent crossover occurs, disagreement is visible. When $m\sim 9.5T$, this pole collides with another purely imaginary pole and its real part begins to grow, in stark disagreement with the coherent hydrodynamic model.}
\label{fig:dispersionrelationtransversecoherent}
\end{figure}

The transverse momentum conductivity $\kappa_\perp\left(\omega,k\right)$, given by the Kubo formula in equation (\ref{eq:transversemomentumconductivitydefninhydro}), determines the response of the momentum (or equivalently heat current) to a source that is modulated in the transverse spatial direction to that of the momentum flow. We find qualitatively different behaviours in the coherent and incoherent regimes, as shown in figure \ref{fig:dispersionrelationtransversecoherent}, in agreement with the analysis of section \ref{sec:hydrotransversesection}.
In the coherent regime $m\ll T$, there is a purely imaginary pole due to a pseudo-diffusive excitation with dispersion relation (\ref{eq:transversemomentumdisprelhydro}), with $\Gamma=m^2/4\pi T$. As $m$ becomes larger than $T$, we approach the coherent/incoherent crossover and the dispersion relation of this pole begins to deviate from that of the hydrodynamic model (\ref{eq:transversemomentumdisprelhydro}). Eventually (at $m\sim9.5T$), the pole collides with another purely imaginary pole and turns into two poles with non-zero real parts, representing the existence of propagating collective excitations. At this point, we are firmly in the incoherent regime: there is no remnant of momentum conservation left in the excitation spectrum as the pseudo-diffusive mode has disappeared.

\section{Self-duality and emergent \texorpdfstring{SL(2,$\mathbb{R}$)$\times$SL(2,$\mathbb{R}$)}{SL(2,R)xSL(2,R)} symmetry \label{section:SelfDuality}}

In the holographic model, there is a special value of the translational symmetry breaking parameter $m=\sqrt{2}\,r_0$ at which the AC heat conductivity is remarkably simple. It is a frequency-independent constant $\kappa(\omega)=\kappa_{DC}$ (and so this point lies in the incoherent regime), which we argue follows from an emergent self-duality at this point, by analogy with Maxwell theory in four dimensions. Precisely at this point, the energy density of the state is zero, as are $\langle T^{xx}\rangle$ and $\langle T^{yy}\rangle$. Moreover, for this value of $m$ there is a further enhancement of the symmetries of the system. Specifically, the linearised perturbations of the gravitational theory exhibit SL(2,$\mathbb{R}$)$\times$SL(2,$\mathbb{R}$) symmetries, which allow us to determine the exact two-point retarded Green's functions of all operators in the dual field theory.

\subsection{Self-duality in Maxwell's equations \label{section:SDMaxwell}}

We begin by briefly summarising the well-known electric/magnetic duality of Maxwell's equations in AdS$_4$ \cite{Witten:2003ya}, and its implications for electrical transport in the dual field theory \cite{Herzog:2007ij}. Consider the action obtained by supplementing our gravitational action (\ref{eq:axiontheoryaction}) with the Maxwell action for a $U(1)$ gauge field
\begin{equation}
\label{action41}
S=\int d^4x \sqrt{-g}\left(\mathcal{R}+6-\frac{1}{2}\sum_{i=1}^{2}\partial^\mu\phi_i\partial_\mu\phi_i-\frac{1}{4}F_{\mu\nu}F^{\mu\nu}\right).
\end{equation}
The Maxwell equation for the field strength $F^{\mu\nu}$, along with the Bianchi identity, imply that the Hodge-dual field strength $G^{\mu\nu}\equiv\frac{1}{2}\epsilon^{\mu\nu\alpha\beta}F_{\alpha\beta}$ satisfies an identical Maxwell equation and Bianchi identity. For a background solution with $F^{\mu\nu}$=0 and with a diagonal metric, such as the Schwarzschild-AdS$_4$ black brane or the translational symmetry-breaking solution (\ref{eq:axiontheorymetricsolution}), this implies that the linear perturbations $F_{rx}\sim \partial_ra_x(r,t,x)$ and $G_{rx}\sim F^{ty}\sim\partial_ta_y(r,t,x)$ obey identical equations of motion (the tildes denote that we are neglecting functions of $r$ for now). This is \textit{self}-duality: the duality transformation relates the fluctuation equations for different fields around the \textit{same} background. Using the holographic dictionary, this relation between the equations of motion for $a_x$ and $a_y$ results in simple relations between the two-point retarded Greens functions of the dual $U(1)$ current operators $J^x$ and $J^y$ (and $J^t$), as we will now show.

The Maxwell equations for the linear perturbations of the gauge field around the translational symmetry-breaking background (\ref{eq:axiontheorymetricsolution}) are
\begin{align}
0&=\frac{d}{dr}\left[r^2\left(a_t'+i\omega a_r\right)\right]-\frac{k}{r^2f}\left(ka_t+\omega a_x\right), \label{eq:gaugeeomdynamic1}\\
0&=\frac{d}{dr}\left[r^2f\left(a_x'-ika_r\right)\right]+\frac{\omega}{r^2f}\left(ka_t+\omega a_x\right), \label{eq:gaugeeomdynamic2}\\
0&=i\omega a_t'+ikfa_x'-\left(\omega^2-k^2f\right)a_r, \label{eq:gaugeeomconstraint}\\
0&=\frac{d}{dr}\left[r^2fa_y'\right]+\frac{\omega^2-k^2f}{r^2f}a_y, \label{eq:gaugeeomdynamic3}
\end{align}
where we have again chosen co-ordinates such that the wavenumber $k$ points in the $x$-direction of the dual field theory, and where primes denote derivatives with respect to $r$. These decouple from the equations for the metric and scalar field fluctuations in this theory, which are given in appendix \ref{sec:numericalappendix}, because the solution is neutral. Due to bulk $U(1)$ gauge-invariance, there are only really two independent dynamical equations, which we take to be (\ref{eq:gaugeeomdynamic2}) and (\ref{eq:gaugeeomdynamic3}). This is because, by combining with the constraint equation (\ref{eq:gaugeeomconstraint}), equation (\ref{eq:gaugeeomdynamic2}) implies the other dynamical equation of motion (\ref{eq:gaugeeomdynamic1}). Of the two remaining degrees of freedom, one corresponds to longitudinal current flow ($a_x\sim J^x$) and the second to transverse current flow ($a_y\sim J^y$).

This reduction, due to symmetry, of the number of independent degrees of freedom has a parallel in the dual field theory. A bulk $U(1)$ gauge field is dual to a conserved $U(1)$ current in the dual field theory, which has an associated Ward identity $\partial_\mu J^\mu=0$. This Ward identity relates the two-point functions of $J^t$ to those of $J^x$ such that there are only two independent quantities controlling all of the retarded Greens functions of the current. We take these to be $G^R_{J^xJ^x}\left(\omega,k\right)$ and $G^R_{J^yJ^y}\left(\omega,k\right)$, with the other non-vanishing Greens functions given by
\begin{equation}
\label{eq:gaugefieldwardidentities}
G^R_{J^xJ^t}\left(\omega,k\right)=G^R_{J^tJ^x}\left(\omega,k\right)=\frac{k}{\omega}G^R_{J^xJ^x}\left(\omega,k\right),\qquad  G^R_{J^tJ^t}=\frac{k^2}{\omega^2}G^R_{J^xJ^x}\left(\omega,k\right),
\end{equation}
up to contact terms. The number of independent field theory two-point Green's functions matches the number of independent dynamical degrees of freedom in the linearised gravitational theory.

To determine the precise relationship between the Green's functions and the solutions to the gravitational equations of motion, we expand the gauge field perturbations near the AdS boundary $a_\mu=a_\mu^{(0)}+a_\mu^{(1)}/r+O(1/r^2)$ and then compute the on-shell action
\begin{equation}
S=\frac{1}{2}\int_{r\rightarrow\infty}\frac{d\omega dk}{4\pi^2}\left[\frac{1}{\omega^2-k^2}\left(ka_t^{(0)}+\omega a_x^{(0)}\right)\left(ka_t^{(1)}+\omega a_x^{(1)}\right)+a_y^{(0)}a_y^{(1)}\right],
\end{equation}
where the first term in each quadratic pair has argument $(-\omega,-k)$ and the second has $(\omega,k)$. Application of the usual holographic procedure then yields
\begin{equation}
\label{eq:expressionsforcurrenttwopointfunctions}
G^R_{J^xJ^x}=-\frac{\omega}{\omega^2-k^2}\frac{\delta\left(ka_t^{(1)}+\omega a_x^{(1)}\right)}{\delta a_x^{(0)}}\Biggr|_{a_t^{(0)}=a_y^{(0)}=0},\qquad G^R_{J^yJ^y}=-\frac{\delta a_y^{(1)}}{\delta a_y^{(0)}}\Biggr|_{a_t^{(0)}=a_x^{(0)}=0},
\end{equation}
along with the Ward identities in equation (\ref{eq:gaugefieldwardidentities}). The gravitational quantities on the right hand sides are determined by solving the two independent dynamical gravitational equations of motion and reading off the appropriate near-boundary quantities.

The above is true in general, and does not rely on electric/magnetic self-duality. The self-duality implies a simple relation between the two independent fluctuation equations, as alluded to above. To make this manifest, it is convenient to work with the bulk ($U(1)$-invariant) variables
\begin{equation}
\label{eq:newbulkgaugefields}
\varphi_{a_x}\equiv r^2f\left(a_x'-ika_r\right),\;\;\;\;\;\;\;\;\;\;\;\varphi_{a_y}\equiv a_y,
\end{equation}
which both obey an \textit{identical} equation of motion
\begin{equation}
\label{eq:gaugefieldselfdualityeom}
\frac{d}{dr}\left(r^2f\varphi'\right)+\frac{\omega^2-k^2f}{r^2f}\varphi=0. 
\end{equation}
Thus, self-duality implies that there is only really $\textit{one}$ independent quantity controlling the $J^\mu$ Green's functions in the dual field theory, even though one would expect there to generically be two such quantities. It implies that both longitudinal ($J^x$) and transverse ($J^y$) current flow are not independent, but are both controlled by a single quantity.

Expanding the new fields (\ref{eq:newbulkgaugefields}) near the boundary, one finds that
\begin{align}
\label{eq:nearboundaryexpansionsgaugefields}
\varphi_{a_x}&=\frac{\omega}{k^2-\omega^2}\left(ka_t^{(1)}+\omega a_x^{(1)}\right)+\frac{\omega}{r}\left(ka_t^{(0)}+\omega a_x^{(0)}\right)+O(1/r^2),\\
\varphi_{a_y}&=a_y^{(0)}+a_y^{(1)}/r+O(1/r^2),
\end{align}
and so, using equation (\ref{eq:expressionsforcurrenttwopointfunctions}), one finds that the Greens functions are given by
\begin{equation}
\label{eq:currentgreensfunctionsfromgifields}
G_{J^xJ^x}\left(\omega,k\right)=\omega^2\frac{\varphi^{(0)}}{\varphi^{(1)}},\qquad G_{J^yJ^y}\left(\omega,k\right)=-\frac{\varphi^{(1)}}{\varphi^{(0)}},
\end{equation}
where $\varphi^{(i)}$ denote the coefficients in the near-boundary expansion $\varphi=\varphi^{(0)}+\varphi^{(1)}/r+O(1/r^2)$. Therefore the self-duality of Maxwell's equations, which manifests itself in the identical equations of motion obeyed by the longitudinal ($\varphi_{a_x}$) and transverse ($\varphi_{a_y}$) components of the gauge field, results in the relation
\begin{equation}
\label{eq:selfdualityinverseproportionalityofcurrents}
G^R_{J^xJ^x}\left(\omega,k\right)G^R_{J^yJ^y}\left(\omega,k\right)=-\omega^2.
\end{equation}
The Green's functions of the longitudinal and transverse components of the conserved $U(1)$ current are inversely proportional to one another. The inverse proportionality, rather than the proportionality that one might expect due to the identical equation of motion, arises because of the qualitative difference in the near-boundary expansions (\ref{eq:nearboundaryexpansionsgaugefields}). One result of this is that the poles of one Green's function correspond to the zeroes of the other.

The inverse proportionality relation (\ref{eq:selfdualityinverseproportionalityofcurrents}) between the Green's functions of the currents results in a constant optical electrical conductivity $\sigma(\omega)=\sigma_{DC}$. In the limit $k=0$, rotational symmetry in the $(x,y)$ plane is restored and thus $G^R_{J^xJ^x}\left(\omega,0\right)=G^R_{J^yJ^y}\left(\omega,0\right)$. Inserting this into (\ref{eq:selfdualityinverseproportionalityofcurrents}) yields $G^R_{J^xJ^x}\left(\omega,0\right)=\pm i\omega$. An application of the Kubo formula then yields
\begin{equation}
\label{eq:electroconductivityconstant}
\sigma\left(\omega\right)=\frac{i}{\omega}\left[G^R_{J^xJ^x}\left(\omega,0\right)-G^R_{J^xJ^x}\left(0,k\rightarrow0\right)\right]=1,
\end{equation}
where we have chosen the sign of $G^R_{J^xJ^x}$ corresponding to a stable state, and where we have assumed that the subtracted term in the middle expression is zero. This assumption is commonly made in the literature, and we will verify it in section \ref{sec:kdependenceselfdualpoint} for the case we are primarily interested in.

This result (\ref{eq:electroconductivityconstant}) for the optical electrical conductivity is \textit{exact}, and is made possible by the self-duality symmetry of Maxwell's equations. This result is well-understood and the corresponding symmetry is relatively common in holographic theories at zero density. For example, it is present for all values of $m$ in the solution (\ref{eq:axiontheorymetricsolution}), including $m=0$ (i.e. the AdS$_4$-Schwarzschild black brane solution). In the following sections we will show that, when $m=\sqrt{2}\,r_0$, the metric and scalar field perturbations also exhibit a self-duality symmetry that produces a constant \textit{heat} conductivity. Furthermore, at this self-dual point, the equations of motion for the perturbations have an emergent SL(2,$\mathbb{R}$)$\times$SL(2,$\mathbb{R}$) symmetry which allows them to be solved exactly, yielding analytic expressions for the full, spatially-resolved Greens functions, and not just their $k=0$ limits as was the case here.  
 
\subsection{Self-duality in the gravitational equations}

We will proceed as we did in the previous section for the case of the gauge field perturbations. There are 12 equations of motion for the coupled metric and scalar perturbations (see appendix \ref{sec:numericalappendix}), but these describe only four dynamical degrees of freedom. All of the two-point retarded Green's functions of the field theory state can be expressed in terms of just four quantities, which are determined by solving the gravitational perturbation equations. The relations between these Green's functions arise from field theory Ward identities, as explained in appendix \ref{sec:selfdualappendix}.

It is convenient to change the operator basis that we use, from $\left\{T^{\mu\nu},\mathcal{O}_i\right\}$ to $\left\{T^{\mu\nu},\Phi_i\right\}$ where $\mathcal{O}_i$ are the operators dual to the gravitational field perturbation $\delta\phi_i$, and the new operators are (in Fourier space) 
\begin{align}
\Phi_1&\equiv\sqrt{2}\,r_0\left(T^{xx}-T^{yy}\right)+2ik\mathcal{O}_1,\\
\Phi_2&\equiv \sqrt{2}\,r_0T^{xy}+ik\mathcal{O}_2.
\end{align}
Having done this, we can write all of the two-point Green's functions in terms of the four following quantities: $G^R_{P^xP^x}\left(\omega,k\right)\equiv G^R_{T^{tx}T^{tx}}\left(\omega,k\right)$, $G^R_{P^yP^y}\left(\omega,k\right)\equiv G^R_{T^{ty}T^{ty}}\left(\omega,k\right)$, $G^R_{\Phi_1\Phi_1}\left(\omega,k\right)$ and $G^R_{\Phi_2\Phi_2}\left(\omega,k\right)$. There is a corresponding convenient basis of (gauge-invariant) gravitational field perturbations, in which the four independent perturbations decouple:
\begin{align}
\Psi_x&=\sqrt{2}\,r_0\left(h^x_x-h^y_y\right)-2ik\delta\phi_1,\label{eq:psixdefn}\\
\varphi_{p_x}&=r\Psi_x-\frac{\sqrt{2}r^4f}{r_0}\left({h^x_x}'+{h^y_y}'-\frac{2}{r}h^r_r-2ikh^x_r-\frac{1}{r^3f}\left(k^2+2r_0^2\right)h^y_y\right),\label{eq:varphipxdefn}\\
\Psi_y&=\sqrt{2}\,r_0h^y_x-ik\delta\phi_2,\label{eq:psiydefn}\\
\varphi_{p_y}&=r^3\left({h^y_t}'+i\omega h^y_r\right),\label{eq:varphipydefn}
\end{align}
where $h_{\mu\nu}\equiv\delta g_{\mu\nu}$. The first two of these control the longitudinal Green's functions in the dual field theory, while the last two control the transverse Green's functions. For conciseness, we do not list the full equations of motion of these fields here. The important point is that, at the special self-dual point $m=\sqrt{2}\,r_0$, the equations of motion of the longitudinal and transverse fields become identical. Both $\varphi_{p_x}$ and $\varphi_{p_y}$ obey the equation
\begin{equation}
\label{eq:firstmetricselfdualeqn}
\frac{d}{dr}\left(r^2f\varphi'\right)+\frac{\omega^2-k^2f}{r^2f}\varphi=0,
\end{equation}
while $\Psi_x$ and $\Psi_y$ both obey
\begin{equation}
\label{eq:secondmetricselfdualeqn}
\frac{d}{dr}\left(r^4f\Psi'\right)+\frac{\omega^2-\left(k^2+2r_0^2\right)f}{f}\Psi=0.
\end{equation}
This is precisely analogous to the properties of the Maxwell equation in the previous section, which could be understood in a very simple way from electric/magnetic self-duality of the field equations. We therefore refer to this as a self-duality of the linearised metric and scalar field equations, although we do not yet know the deeper underlying reason for the symmetry in this case. Note that the equation of motion (\ref{eq:firstmetricselfdualeqn}) is identical to the Maxwell gauge field equation of motion (\ref{eq:gaugefieldselfdualityeom}) when $m=\sqrt{2}\,r_0$: the fields $\varphi_{p_x}$ and $\varphi_{p_y}$ literally obey the Maxwell equations at the self-dual point. Gravitational self-dualities of this type have previously been identified in pure AdS space \cite{Leigh:2003ez,Julia:2005ze,Leigh:2007wf,deHaro:2008gp,Bakas:2008gz,Bakas:2008zg}, as well as in ``partially massless'' gravity in dS space \cite{Deser:2013xb,Hinterbichler:2014xga}. Given the close similarities between our axionic theory and massive gravity \cite{Andrade:2013gsa,Taylor:2014tka}, as well as the SL(2,$\mathbb{R}$) symmetry of our axionic theory at this self-dual point (see below) and the corresponding symmetry of the wave equation in dS \cite{Anninos:2011af}, it would be very interesting to determine the relation between our theory and the partially massless theory more precisely.

This symmetry therefore results in a reduction in the number of independent quantities that determine all of the Green's functions. Although normally there are 4 such quantities in the metric and scalar sector of the theory, this is reduced to 2 at the self-dual point: the solutions to the two equations of motion (\ref{eq:firstmetricselfdualeqn}) and (\ref{eq:secondmetricselfdualeqn}). To precisely extract the Greens functions from the solutions to these equations of motion, a careful analysis of the on-shell action must be performed. This is presented in appendix \ref{sec:selfdualappendix} and here we will just give the results:
\begin{align}
G^R_{P^xP^x}\left(\omega,k\right)&=\omega^2\left(k^2+2r_0^2\right)\frac{\varphi^{(0)}}{\varphi^{(1)}},\label{eq:selfdualcorrelator1intermsofbuildingblock}\\
G^R_{P^yP^y}\left(\omega,k\right)&=-\left(k^2+2r_0^2\right)\frac{\varphi^{(1)}}{\varphi^{(0)}},\label{eq:selfdualcorrelator2intermsofbuildingblock}\\
G^R_{\Phi_1\Phi_1}\left(\omega,k\right)&=-12{\left(k^2+2r_0^2\right)}\frac{\Psi^{(3)}}{\Psi^{(0)}},\label{eq:selfdualcorrelator3intermsofbuildingblock}\\
G^R_{\Phi_2\Phi_2}\left(\omega,k\right)&=-3\left(k^2+2r_0^2\right)\frac{\Psi^{(3)}}{\Psi^{(0)}},\label{eq:selfdualcorrelator4intermsofbuildingblock}
\end{align}
where $\varphi^{(i)}$ and $\Psi^{(i)}$ denote the coefficients of the $r^{-i}$ term in the near-boundary expansions of $\varphi$ and $\Psi$ respectively.

The most important thing to note is that the self-duality symmetry relating the longitudinal and transverse degrees of freedom in the gravitational theory ensures that the two-point retarded Green's functions of the transverse and longitudinal momentum currents (or, equivalently, heat currents) are inversely proportional to one another
\begin{equation}
\label{eq:inverserelationshipselfdualmomentum}
G^R_{P^xP^x}\left(\omega,k\right)G^R_{P^yP^y}\left(\omega,k\right)=-\omega^2\left(k^2+2r_0^2\right)^2,
\end{equation}
so that the poles of one Green's functions will correspond to the zeroes of the other. This is analogous to the gauge field case in the previous section, where the corresponding quantities were the transverse and longitudinal $U(1)$ currents. Using the presence of rotational symmetry in the $k=0$ limit, we find that $G^R_{P^xP^x}\left(\omega,0\right)^2=-4r_0^4\omega^2$ and therefore that the AC heat conductivity $\kappa(\omega)$ is exactly frequency independent at the self-dual point $m=\sqrt{2}\,r_0$
\begin{equation}
\label{eq:selfdualheatcondfromkubo}
\kappa\left(\omega\right)=\frac{i}{\omega T}\left[G^R_{P^xP^x}\left(\omega,0\right)-G^R_{P^xP^x}\left(0,k\rightarrow0\right)\right]=\frac{2r_0^2}{T}=8\pi^2 T.
\end{equation}
We have again assumed here that the subtracted term in the second equation above vanishes, and will prove this in the following section. Note that this is consistent with the analytic result (\ref{eq:analytickappadcequation}) for $\kappa_{DC}$.

In summary, the equations of motion for the perturbations of the metric and scalar fields exhibit a self-duality symmetry at the special value $m=\sqrt{2}\,r_0$, in that the fields transverse to the wavenumber obey identical equations of motion as those parallel to it. The result of this is that all two point retarded Green's functions (of $T^{\mu\nu}$, $J^\mu$ and $\Phi_i$) are determined by just two independent quantities, instead of six as one would generically expect. The self-duality implies that the AC heat conductivity $\kappa(\omega)$ is exactly frequency independent (like the optical electrical conductivity $\sigma(\omega)$), and that its value (\ref{eq:selfdualheatcondfromkubo}) can be determined without having to explicitly solve the gravitational equations of motion. 

\subsection{Emergent \texorpdfstring{SL(2,$\mathbb{R}$)$\times$SL(2,$\mathbb{R}$)}{SL(2,R)xSL(2,R)} symmetry: Exact Green's functions}
\label{sec:kdependenceselfdualpoint}

We have just described the emergent self-duality symmetry present when $m=\sqrt{2}\,r_0$, which reduces the number of independent Green's functions in the dual field theory to just two. There is a second, independent symmetry enhancement which occurs at this special self-dual value of $m$, which allows us to determine these quantities, and hence all two-point retarded Green's functions of the theory \eqref{action41}, \textit{exactly}, for all $\omega$ and $k$. This is highly unusual: in almost every holographic theory, Green's functions can only be computed approximately (either numerically or analytically in certain limits). Even for very simple cases like the two-point functions of $T^{\mu\nu}$ or $J^\mu$ in the neutral, thermal state dual to the AdS-Schwarzschild black brane solution, exact results are not available. Yet in this seemingly complicated setup, in which translational symmetry is explicitly broken, the two-point Green's functions of all operators (including the current operator of section \ref{section:SDMaxwell}) can be determined exactly. 

All two-point Green's functions are controlled by the solutions to the equations of motion \eqref{eq:gaugefieldselfdualityeom}, (\ref{eq:firstmetricselfdualeqn}) and (\ref{eq:secondmetricselfdualeqn}). These equations can be solved analytically: their general solutions are linear combinations of hypergeometric functions $_2F_1$. After imposing ingoing boundary conditions at the horizon, they are
\begin{align}
\varphi=&\varphi^{(0)}\left(1-\frac{r_0^2}{r^2}\right)^{-\frac{i\omega}{2r_0}}\Biggl\{\;_2F_1\left[\frac{1}{4}-\frac{i\omega}{2r_0}+\frac{\nu_1}{2},\frac{1}{4}-\frac{i\omega}{2r_0}-\frac{\nu_1}{2},\frac{1}{2},\frac{r_0^2}{r^2}\right]\\
&-\frac{2r_0}{r}\frac{\Gamma\left(\frac{3}{4}-\frac{i\omega}{2r_0}-\frac{\nu_1}{2}\right)\Gamma\left(\frac{3}{4}-\frac{i\omega}{2r_0}+\frac{\nu_1}{2}\right)}{\Gamma\left(\frac{1}{4}-\frac{i\omega}{2r_0}-\frac{\nu_1}{2}\right)\Gamma\left(\frac{1}{4}-\frac{i\omega}{2r_0}+\frac{\nu_1}{2}\right)}\;_2F_1\left[\frac{3}{4}-\frac{i\omega}{2r_0}-\frac{\nu_1}{2},\frac{3}{4}-\frac{i\omega}{2r_0}+\frac{\nu_1}{2},\frac{3}{2},\frac{r_0^2}{r^2}\right]\Biggr\},\nonumber\\
\Psi=&\Psi^{(0)}\left(1-\frac{r_0^2}{r^2}\right)^{-\frac{i\omega}{2r_0}}\Biggl\{\;_2F_1\left[-\frac{1}{4}-\frac{i\omega}{2r_0}+\frac{\nu_2}{2},-\frac{1}{4}-\frac{i\omega}{2r_0}-\frac{\nu_2}{2},-\frac{1}{2},\frac{r_0^2}{r^2}\right]\\
&+\frac{8}{3}\frac{r_0^3}{r^3}\frac{\Gamma\left(\frac{5}{4}-\frac{i\omega}{2r_0}+\frac{\nu_2}{2}\right)\Gamma\left(\frac{5}{4}-\frac{i\omega}{2r_0}-\frac{\nu_2}{2}\right)}{\Gamma\left(-\frac{1}{4}-\frac{i\omega}{2r_0}+\frac{\nu_2}{2}\right)\Gamma\left(-\frac{1}{4}-\frac{i\omega}{2r_0}-\frac{\nu_2}{2}\right)}\;_2F_1\left[\frac{5}{4}-\frac{i\omega}{2r_0}+\frac{\nu_2}{2},\frac{5}{4}-\frac{i\omega}{2r_0}-\frac{\nu_2}{2},\frac{5}{2},\frac{r_0^2}{r^2}\right]\Biggr\}\nonumber,
\end{align}
where $\nu_1=\sqrt{1/4-k^2/r_0^2}$ and $\nu_2=i\sqrt{7/4+k^2/r_0^2}$. From these, it is easy to read off the values of the two independent quantities which control all of the two-point Green's functions of the theory
\begin{align}
\mathcal{G}_A\left(\omega,k\right)&\equiv\frac{\varphi^{(1)}}{\varphi^{(0)}}=-2r_0\frac{\Gamma\left(\frac{3}{4}-\frac{i\omega}{2r_0}-\frac{1}{4}\sqrt{1-4\frac{k^2}{r_0^2}}\right)\Gamma\left(\frac{3}{4}-\frac{i\omega}{2r_0}+\frac{1}{4}\sqrt{1-4\frac{k^2}{r_0^2}}\right)}{\Gamma\left(\frac{1}{4}-\frac{i\omega}{2r_0}-\frac{1}{4}\sqrt{1-4\frac{k^2}{r_0^2}}\right)\Gamma\left(\frac{1}{4}-\frac{i\omega}{2r_0}+\frac{1}{4}\sqrt{1-4\frac{k^2}{r_0^2}}\right)},\label{eq:selfdualbuildingblock1}\\
\mathcal{G}_B\left(\omega,k\right)&\equiv\frac{\Psi^{(3)}}{\Psi^{(0)}}=\frac{8r_0^3}{3}\frac{\Gamma\left(\frac{5}{4}-\frac{i\omega}{2r_0}+\frac{i}{4}\sqrt{7+4\frac{k^2}{r_0^2}}\right)\Gamma\left(\frac{5}{4}-\frac{i\omega}{2r_0}-\frac{i}{4}\sqrt{7+4\frac{k^2}{r_0^2}}\right)}{\Gamma\left(-\frac{1}{4}-\frac{i\omega}{2r_0}+\frac{i}{4}\sqrt{7+4\frac{k^2}{r_0^2}}\right)\Gamma\left(-\frac{1}{4}-\frac{i\omega}{2r_0}-\frac{i}{4}\sqrt{7+4\frac{k^2}{r_0^2}}\right)}.\label{eq:selfdualbuildingblock2}
\end{align}

Before inspecting the properties of the exact Green's functions we have obtained, it is instructive to ask why we can solve these equations of motion exactly, at this special value of $m$. The reason is that, at this value, the metric is conformal to a patch of AdS$_2\times R^2$, and thus the perturbation equations have an SL(2,$\mathbb{R}$)$\times$SL(2,$\mathbb{R}$) symmetry. These structures are well-studied in many places, for instance in dS/CFT \cite{Anninos:2011af}, and also in the near-horizon geometries of certain AdS black branes \cite{Anantua:2012nj}, and we refer the interested reader to the former reference for more details. The operators
\begin{align}
a_0&=\left(\frac{\partial}{\partial\theta}+r_0\frac{\partial}{\partial t}\right),\\
a_\pm&=-ie^{\pm\left(\theta+t/r_0\right)}\frac{1}{2}\left(\pm\frac{\partial}{\partial\rho}-\tanh\rho\frac{\partial}{\partial\theta}-r_0\coth\rho\frac{\partial}{\partial t}\right),\\
b_0&=\left(-\frac{\partial}{\partial\theta}+r_0\frac{\partial}{\partial t}\right),\\
b_\pm&=-ie^{\mp\left(\theta-t/r_0\right)}\frac{1}{2}\left(\pm\frac{\partial}{\partial\rho}+\tanh\rho\frac{\partial}{\partial\theta}-r_0\coth\rho\frac{\partial}{\partial t}\right),
\end{align}
are the generators of two commuting SL(2,$\mathbb{R}$) algebras obeying $\left[a_+,a_-\right]=a_0$, $\left[a_0,a_\pm\right]=\pm2 a_\pm$ and $\left[b_+,b_-\right]=b_0$, $\left[b_0,b_\pm\right]=\pm2 b_\pm$. The Casimir of each algebra is given by $C_a=a_0^2/2+a_+a_-+a_-a_+$ and similarly for $C_b$. From these Casimirs, we can build the equations
\begin{align}
\left(C_a+C_b+\frac{3}{4}\right)e^{-i\omega t/r_0^2+ip_\varphi\theta}\frac{\tilde{\varphi}}{\sqrt{\sinh2\rho}}=0,\\
\left(C_a+C_b-\frac{5}{4}\right)e^{-i\omega t/r_0^2+ip_\Psi\theta}\frac{\tilde{\Psi}}{\sqrt{\sinh2\rho}}=0,
\end{align}
where $p_\varphi^2=k^2/r_0^2-1/4$ and $p_\Psi^2=k^2/r_0^2+7/4$. These two equations are equivalent to the original perturbation equations (\ref{eq:firstmetricselfdualeqn}) and (\ref{eq:secondmetricselfdualeqn}) after changing co-ordinates to $\rho=\log\coth\left(\frac{1}{2}\tanh^{-1}\frac{r_0}{r}\right)$ and rescaling the fields $\tilde{\varphi}=\sqrt{\sinh\rho}\varphi$ and $\tilde{\Psi}=r\sqrt{\sinh\rho}\Psi$.

The exact two-point retarded Green's functions of all operators in the theory are given by combining the expressions (\ref{eq:gaugefieldwardidentities}),  (\ref{eq:currentgreensfunctionsfromgifields}), (\ref{eq:wardidentitieslist}) and equations (\ref{eq:selfdualcorrelator1intermsofbuildingblock}) to (\ref{eq:selfdualcorrelator4intermsofbuildingblock}) with the solutions to the equations of motion given in equations (\ref{eq:selfdualbuildingblock1}) and (\ref{eq:selfdualbuildingblock2}). In particular, the longitudinal momentum and current Green's functions are
\begin{align}
G^R_{P^xP^x}\left(\omega,k\right)=&-\frac{\omega^2\left(k^2+2r_0^2\right)}{2r_0}\frac{\Gamma\left(\frac{1}{4}-\frac{i\omega}{2r_0}-\frac{1}{4}\sqrt{1-4\frac{k^2}{r_0^2}}\right)\Gamma\left(\frac{1}{4}-\frac{i\omega}{2r_0}+\frac{1}{4}\sqrt{1-4\frac{k^2}{r_0^2}}\right)}{\Gamma\left(\frac{3}{4}-\frac{i\omega}{2r_0}-\frac{1}{4}\sqrt{1-4\frac{k^2}{r_0^2}}\right)\Gamma\left(\frac{3}{4}-\frac{i\omega}{2r_0}+\frac{1}{4}\sqrt{1-4\frac{k^2}{r_0^2}}\right)},\label{eq:pxpxselfdualcorrelator}\\
G^R_{J^xJ^x}\left(\omega,k\right)=&-\frac{\omega^2}{2r_0}\frac{\Gamma\left(\frac{1}{4}-\frac{i\omega}{2r_0}-\frac{1}{4}\sqrt{1-4\frac{k^2}{r_0^2}}\right)\Gamma\left(\frac{1}{4}-\frac{i\omega}{2r_0}+\frac{1}{4}\sqrt{1-4\frac{k^2}{r_0^2}}\right)}{\Gamma\left(\frac{3}{4}-\frac{i\omega}{2r_0}-\frac{1}{4}\sqrt{1-4\frac{k^2}{r_0^2}}\right)\Gamma\left(\frac{3}{4}-\frac{i\omega}{2r_0}+\frac{1}{4}\sqrt{1-4\frac{k^2}{r_0^2}}\right)}.\label{eq:jxjxselfdualcorrelator}
\end{align}
As expected, applications of the Kubo formulae yield the constant results (\ref{eq:electroconductivityconstant}) and (\ref{eq:selfdualheatcondfromkubo}) for the AC conductivities that we previously inferred using self-duality. In particular, the subtracted terms in the Kubo formulae both vanish given the Green's functions (\ref{eq:pxpxselfdualcorrelator}) and (\ref{eq:jxjxselfdualcorrelator}).

Going beyond AC conductivities, the expressions (\ref{eq:pxpxselfdualcorrelator}) and (\ref{eq:jxjxselfdualcorrelator}) contain the full spatial dependence of the Green's functions (and hence the conductivities). Both Green's functions possess poles with the dispersion relation
\begin{equation}
\omega=-ir_0^{-1}k^2+O(k^4),
\end{equation}
which correspond to the existence of diffusive excitations in the theory with diffusion constant $D=r_0^{-1}$. Due to the Ward identities for the conservation of $U(1)$ charge and energy, these are also excitations in the Green's functions of charge density and energy density. These are the only low energy excitations in the theory (i.e. the only excitations with $\omega\rightarrow0$ as $k\rightarrow0$) and simply represent the fact that both charge and energy diffuse, due to the conservation laws of these quantities (recall the discussion in section \ref{section:incoherent}). Generically, the diffusion constants of charge and energy in a theory will be different, although in this case they coincide. The value $D=r_0^{-1}$ for each diffusion constant agrees with the Einstein relations $D=\sigma_{DC}/\chi$ and $D=\kappa_{DC}/(T\partial s/\partial T)$ for the charge and energy diffusion constants respectively, where $\chi$ is the charge susceptibility.\footnote{$\chi$ can be computed from the expressions in \cite{Andrade:2013gsa}, and is equal to $r_0$ for the neutral theory with $m^2=2r_0^2$.} The exact Green's functions we have obtained at this highly symmetric point clearly contain non-trivial excitations, consistent with what one would expect from general principles.

By examining the `building blocks' (\ref{eq:selfdualbuildingblock1}) and (\ref{eq:selfdualbuildingblock2}) from which all Green's functions are constructed, it can be seen that they do not contain any poles (nor zeroes, in the case of $\mathcal{G}_A$) that lie in the upper half of the complex frequency plane and that would indicate perturbative instabilities of the state.

Finally, we can define a `viscosity' for this state using the usual Kubo formula
\begin{equation}
\eta=-\lim_{\omega\rightarrow0}\frac{1}{\omega}\text{Im}G^R_{T^{xy}T^{xy}}\left(\omega,k=0\right)\approx0.325\times\frac{s}{4\pi}.
\end{equation}
This violates the KSS viscosity bound \cite{Kovtun:2004de} (it avoids the general proof \cite{Iqbal:2008by} because, due to the translational symmetry breaking, the equation of motion for $h^x_y$ is no longer that of a minimally coupled massless scalar field). However, this is not a conventional viscosity controlling, for example, the rate of transverse momentum diffusion, since our system is no longer described by hydrodynamics in this incoherent regime. A violation of the viscosity bound was found in a related holographic model in \cite{Jain:2014vka}.

\section{Charge transport in non-Maxwell theories \label{section:NonMaxwell}}

In this section, we will address some holographic models of strongly interacting field theory states with a global $U(1)$ symmetry. We will primarily be interested in their electrical conductivity: the linear response of the $U(1)$ current $\vec{J}$ to an external source. In a strongly interacting system of this kind, there is no reason to expect that the current is almost-conserved (there is no symmetry protecting it) and thus one would expect the charge transport to be incoherent. Coherent charge transport arises naturally in holographic models in which the current overlaps with the momentum, which can be almost-conserved \cite{Hartnoll:2012rj}. Here, we will instead focus on states in which the current and momentum are \textit{not} coupled, i.e.~perturbations of the gauge field in the gravitational theory decouple from perturbations of the metric. We find that there \textit{are} coherent regimes of charge transport, despite the fact there is no obvious symmetry producing an almost-conserved quantity. 

The two kinds of holographic theory we study -- probe brane systems and neutral systems with non-minimal gauge field actions -- are deformations of the Maxwell action. The Maxwell theory is incoherent: it has a constant optical conductivity $\sigma\left(\omega\right)=\sigma_{DC}$, and charge transport occurs via diffusion \cite{Herzog:2007ij}. Upon deforming this theory -- with a probe brane charge density or a higher-derivative gauge field term in the action -- coherent peaks of the form (\ref{eq:drudepeakintroduction}) can emerge. 

\subsection{Probe brane theories at non-zero density}

Firstly, we will address simple probe brane theories with a trivial embedding, for which the action may schematically be written
\begin{equation}
\label{eq:generalprobebraneaction}
S=N_c^2\int d^{d+2}x\sqrt{-g}\left(\mathcal{R}+\ldots\right)+N_c\int d^{p+1}x\sqrt{-\text{det}\left(g_{ab}+F_{ab}\right)},
\end{equation}
where $N_c\rightarrow\infty$ and $\left\{a,b\right\}$ go from $0\ldots p$. These theories have a long and rich history and we refer the reader to \cite{Karch:2007pd,Karch:2007br,Hartnoll:2009ns} and their references for more details. The $O(N_c^2)$ terms, which dominate in this limit, determine the metric of the background solution. The subleading $O(N_c)$ term is important because it controls the charge transport properties of the theory: it is the leading order at which the gauge field $F_{ab}$ appears. Physically, one can think of this as a system with $O(N_c)$ charged degrees of freedom living in $p-1$ spatial dimensions coupled to a bath of $O(N_c^2)$ neutral degrees of freedom living in $d$ spatial dimensions. In this probe limit, linear perturbations of the metric and the gauge field decouple. This means that the cross correlator of $\vec{J}$ with $\vec{P}$ vanishes in the dual field theory state: the dynamics of the current do not affect the dynamics of the momentum since the contribution of the charged degrees of freedom to $\vec{P}$ is suppressed by $1/N_c$ with respect to the contribution of the neutral degrees of freedom.

In a zero density state (i.e. when the background solution has $A_t=0$), the gauge field perturbations obey the Maxwell equations and thus there is incoherent transport and charge diffusion. At non-zero densities ($A_t\ne0$), non-linear effects in the action (\ref{eq:generalprobebraneaction}) become important and as the ratio of the $U(1)$ chemical potential to the temperature $\mu/T$ increases, these effects are more pronounced and qualitatively change the system's electrical transport properties. For a probe brane in a geometry corresponding to a neutral sector with dynamical critical exponent $z$ and hyperscaling violation exponent $\theta$, there is a delta function in the $T/\mu=0$ optical conductivity when \cite{HoyosBadajoz:2010kd,Pang:2013ypa,Dey:2013vja,Edalati:2013tma}
\begin{equation}
\label{eq:probebraneinequality}
z<2\left(1-\theta/d\right),
\end{equation}
assuming that $z>0$ and $\theta\leq0$. For $z$ larger than this upper bound, there is no delta function the $T=0$ optical conductivity, which now grows as a power law at small $\omega$.

In systems that obey the inequality (\ref{eq:probebraneinequality}), there is a very clear agreement with the effective theory of coherent transport outlined in the introduction. For simplicity we will focus on the case $z=1$, $\theta=0$ (e.g. the supersymmetric D3/D5 and D3/D7 systems) which has been studied the most thoroughly. The characteristic dissipation rate in this state is $\Gamma\sim T^2/\mu$, and may be read off from the coherent peak observed in the non-zero temperature optical conductivity \cite{Hartnoll:2009ns}. This is consistent with the spatially resolved charge transport: at very long distances $k\ll\Gamma$, charge is transported via diffusion while at shorter distances $\Gamma\ll k$, the longest lived excitation that carries charge is the ``holographic zero sound'' mode \cite{Karch:2008fa,Bergman:2011rf,Davison:2011ek}. A similar transition at intermediate distance scales is found in the Sakai-Sugimoto model, where $\Gamma\sim T^3/\mu^2$ \cite{Kulaxizi:2008jx,DiNunno:2014bxa}. 

A simple explanation of this phenomenology would be that the system has an almost-conserved quantity, with dissipation rate $\Gamma\sim T^2/\mu$, that overlaps with the current. This would-be explanation begs the questions of what is conserved when $T=0$ and why? In our hydrodynamic model of heat conductivity in section \ref{section:HeatTransportGen}, the conserved quantity is the total momentum, and it is conserved due to a symmetry: translational invariance. This also makes it clear that $\Gamma$ in that case should increase with $m$ -- the parameter controlling the translational symmetry breaking -- and that energy should be transported at short distances by sound-like waves -- the excitations characteristic of a system with both conserved momentum and energy. One possible answer for our probe brane theories is that the current itself becomes conserved in the $T=0$ limit, but it is not clear why this would be the case in a strongly interacting system, why it should depend on $T$, and why the characteristic excitation associated with conservation of both charge and current would be the holographic zero sound mode. We note that the $z=1$, $\theta=0$ solution has a boost symmetry in the limit $T=0$ \cite{Karch:2008uy}, and it would be interesting to see if this generalises to other values of $z,\theta$ satisfying the inequality (\ref{eq:probebraneinequality}). On a related note, the existence of zero temperature diffusive excitations in these states was found to be a result of the existence of moduli spaces for certain scalar operators \cite{Ammon:2012mu}. It could be the case that the probe limit is singular and it would also be worthwhile to go beyond this limit in these kind of systems (e.g. \cite{Tarrio:2013tta}): this may qualitatively change the low energy transport when even a small amount of backreaction is taken into account, as the current will then couple to the total momentum, which is conserved.

For values of $z$ larger than the upper bound in the inequality (\ref{eq:probebraneinequality}), less information is available. At least when $\theta=0$, there appears to be a coherent peak in the optical conductivity at small, non-zero $T/\mu$, for all values of $z$ \cite{Hartnoll:2009ns}. Unlike in the cases satisfying the inequality (\ref{eq:probebraneinequality}), there is no delta function peak in the $T=0$ optical conductivity, but a power law divergence as $\omega\rightarrow0$, indicating that the $\omega\rightarrow0$ and $T\rightarrow0$ limits do not commute. Furthermore, the $k\ne0$ excitation at zero temperature is no longer the holographic zero sound mode \cite{HoyosBadajoz:2010kd,Pang:2013ypa,Dey:2013vja,Edalati:2013tma}. It would be worthwhile exploring these type of states more carefully to determine how exactly they transport charge, but this is beyond the scope of this paper.

\subsection{Neutral theories with higher-derivative couplings}

The second class of systems that we look at are neutral states of theories in which the gauge field action of the gravitational theory includes not only the Maxwell term, but higher-derivative corrections to this arising due to couplings of the gauge field to the Weyl tensor. In particular, consider the action
\begin{equation}
\label{eq:higherderivativegeneralaction}
S=\int d^4x\sqrt{-g}\left(\mathcal{R}+6-\frac{1}{4}F_{\mu\nu}F^{\mu\nu}+\gamma_1 C_{\mu\nu\alpha\beta}F^{\mu\nu}F^{\alpha\beta}-\gamma_2C_{\alpha\beta\rho\sigma}C^{\rho\sigma\alpha\beta}F_{\mu\nu}F^{\mu\nu}\right),
\end{equation}
where $C_{\alpha\beta\rho\sigma}$ is the Weyl tensor. The relevant neutral solution to this action is the neutral AdS$_4$-Schwarzschild black brane. When $\gamma_1=\gamma_2=0$, the action (\ref{eq:higherderivativegeneralaction}) reduces to the Maxwell action and so the optical conductivity $\sigma\left(\omega\right)$ is constant due to self-duality, and charge diffuses at non-zero wavenumbers \cite{Herzog:2007ij}. There is incoherent charge transport. When either of $\gamma_i$ are non-zero, this self-duality is broken \cite{Myers:2010pk,WitczakKrempa:2012gn,Chowdhury:2012km,WitczakKrempa:2013ht,Witczak-Krempa:2013nua,Witczak-Krempa:2013aea,Bai:2013tfa}, and peaks can begin to appear. As in the probe brane examples, we do not know of any reason (such as an approximate conservation law) why this is the case. 

When $\gamma_1>0$ (and $\gamma_2=0$), a low frequency peak appears in the optical conductivity which becomes sharper and taller as $\gamma_1$ is increased. This peak arises due to a pole at $\omega=-i\Gamma$ where $\Gamma\sim T/\gamma_1^{0.66}$, which becomes long-lived with respect to the other excitations of the theory (with decay rates $\Lambda\sim T$) in the limit $\gamma_1\gg1$. Thus there is coherent charge transport in the limit $\gamma_1\gg1$, with a transition to incoherent charge transport when $\gamma_1\sim1$. However, $\gamma_1$ is bounded by causality, and must satisfy $-1/12\leq\gamma_1\leq1/12$ \cite{Hofman:2008ar,Myers:2010pk}. When $\gamma_1<0$, there is no long-lived excitation in the conductivity (in fact, the closest feature to the origin is a zero), and thus the transport is incoherent.

When $\gamma_2>0$ (and $\gamma_1=0$), there are similar results. A low frequency peak appears in the optical conductivity which becomes sharper and taller at larger $\gamma_2$. This is due to a pole in the conductivity at $\omega=-i\Gamma$ where $\Gamma\sim T/\gamma_2^{0.83}$ \cite{Witczak-Krempa:2013aea}, which is well-separated from the other excitations with imaginary parts $\Lambda\sim T$. Again, there is coherent charge transport when $\gamma_2\gg1$, with a transition to incoherent charge transport when $\gamma_2\sim1$. However, unlike in the previous example, the coefficient $\gamma_2$ is \textit{not} bounded from above and so there is a coherent regime of charge transport in this theory. However, this coherent regime exists only when the six-derivative term in the action (\ref{eq:higherderivativegeneralaction}) is more important than the two- and four-derivative terms. This is totally at odds with the effective field theory approach taken in writing down the action (\ref{eq:higherderivativegeneralaction}), and so it would be unwise to take this coherent regime too seriously. The $\gamma_2$ term in the action (\ref{eq:higherderivativegeneralaction}) plays the role of an $r$-dependent coupling constant for the $U(1)$ gauge field, and it would be worthwhile to investigate if the same effect (i.e.~coherent charge transport) could be obtained in a neutral Einstein-Maxwell-dilaton system with exponential coupling, in which the $r$-dependent coupling constant is generated by the dilaton rather than by a higher derivative term (see \cite{Katz:2014rla} for work in this direction). Indeed, it is known that electro-magnetic duality can be restored in this case by also flipping the sign of the coupling between the dilaton and the Maxwell term in the action and field equations \cite{Charmousis:2009xr}.

\section{Discussion \label{section:Discussion}}

In this paper, we have studied how heat is transported in neutral systems in which momentum is not conserved. Firstly, we examined a hydrodynamic system in which the conservation equations were modified to include momentum dissipation. We then compared its transport properties to those of a strongly coupled, holographic system, in which momentum dissipates due to the breaking of translational invariance by linearly dependent scalar sources controlled by a parameter $m$, and found excellent agreement. At sufficiently small $m$, momentum dissipates slowly at the rate $\Gamma\sim m^2/T$. The low energy, spatially uniform transport properties of these systems are controlled by the interplay of two scales: the momentum dissipation rate $\Gamma$, and the thermal scale $\Lambda\sim T$. The systems can be broadly classified into two regimes: the \textit{coherent regime} when $\Gamma\ll\Lambda$, and the \textit{incoherent} regime when $\Gamma\gtrsim\Lambda$. Within each of these regimes, the spatially resolved transport properties depend upon the distance scale one examines.

In the coherent regime, the low frequency AC conductivity at zero $k$ has a Drude-like form (figure \ref{fig:ConductivityNeutralk=0}). The late time transport properties are dominated by a single pole in the complex frequency plane with dispersion relation $\omega=-i\Gamma$, which is parametrically separated from the other poles with decay rates $\sim\Lambda$ (figure \ref{fig:zerokpoles}). At non-zero $k$, the heat transport mechanism depends upon the distance scale. Over the longest distance scales $k\ll\Lambda\ll\Gamma$, momentum significantly dissipates and the heat transport is dominated by diffusion. Over shorter distances $\Lambda\ll k\ll\Gamma$, momentum is long-lived and sound-like modes carry heat. As $k$ is varied, the crossover between these is realised by a collision of poles of the conductivity in the complex frequency plane (figure \ref{fig:ComplexWPlanePolesHolo}).

In the incoherent regime $\Gamma\gtrsim\Lambda$, the low energy AC conductivity no longer has a simple Drude-like form, and there is no longer a Drude-like collective excitation that is well-separated from the other collective excitations of the theory. In our holographic model, we have shown that as $\Gamma$ is increased, the heat conductivity changes from a sharp Drude-like peak to a constant, and then to a valley (figure \ref{fig:ConductivityNeutralk=0}). The longitudinal heat transport is always diffusive in this regime, as momentum is never approximately conserved. Although this is in agreement with the statement of \cite{Hartnoll:2014lpa} that incoherent transport is mediated by diffusive processes, we emphasise that long distance transport in the coherent regime is also mediated by diffusive processes. The difference between them lies at shorter distance scales, where in the coherent case a new excitation, associated with the approximate conservation law, takes over. At low temperatures $T\ll m$ (which coincides with the incoherent regime), the diffusion constant does not depend on temperature, $D_\parallel=\sqrt{3/2}/m$, and the dimensionless combination $D_\parallel T$ can be made arbitrarily small. It would be interesting to re-examine it at non-zero density in light of the bound on energy/charge diffusion constants proposed in \cite{Hartnoll:2014lpa} (see also \cite{Kovtun:2014nsa}).

In the coherent regime, the approximate conservation of momentum means that heat is transported in the transverse directions by a pseudo-diffusive excitation (figure \ref{fig:dispersionrelationtransversecoherent}). At shorter distances than this, and in the incoherent regime, there is no approximate conservation law and thus no long-lived collective excitations.

In this paper, we have focused mainly on the heat conductivity and its poles in the complex frequency plane. There are other potentially important features in the conductivity besides its poles, such as its zeroes \cite{WitczakKrempa:2012gn,WitczakKrempa:2013ht}, and other analytic contributions at low frequencies. It would be interesting to study these more carefully in our model. It would also be interesting to extend our analysis of low energy transport $\omega\ll T$ to study processes at higher energies $T\ll\omega\ll m$. We have focused on a holographic model where the zero temperature IR fixed point is given by an AdS$_2\times R^2$ geometry. It is easy to generalise this geometry to other values of the exponents $z$ and $\theta$, by introducing couplings to a neutral scalar \cite{Charmousis:2010zz,Gouteraux:2011ce,Gouteraux:2014hca}, and it would be interesting to ascertain the effects of this on the transport properties. Including a constant background magnetic field would also help bridge the results of \cite{Hartnoll:2007ih,Hartnoll:2007ip,Hartnoll:2008hs,Blake:2013bqa}.

A more pressing question is to generalise our results to systems with a charge density, such that the system's momentum and current are coupled. The DC conductivities of the two charges have been investigated in holographic systems of this type \cite{Blake:2013bqa, Blake:2013owa, Andrade:2013gsa,Davison:2013txa, Donos:2014uba, Gouteraux:2014hca, Lucas:2014zea,Blake:2014yla, Donos:2014cya, Donos:2014oha, Taylor:2014tka, Amoretti:2014mma, Donos:2014yya} but, as we have demonstrated, this is not sufficient to infer the nature (i.e.~coherent or incoherent) of the conduction of each charge in a system. For very weak translational symmetry breaking, the DC electrical conductivity in these systems is not directly proportional to the momentum dissipation rate, indicating that there is a richer structure in the low energy physics of the system. This is presumably due to the existence of an extra conserved charge (the electrical $U(1)$ charge) and its associated diffusive excitation. Calculations to determine the precise nature of the transport in these systems are underway, and we hope to report on this soon.

Besides this pressing question, there are further interesting questions raised by our results. We found that for a specific value of the parameter $m$ of our holographic model, there is a constant AC heat conductivity due to a self-duality of the gravitational perturbations analogous to that of Maxwell theory in (Schwarzschild-)AdS$_4$. In the Maxwell case, small deviations from the self-dual action turn this self-duality into duality relating two different theories: one with ``particle-like'' excitations carrying charge and one with ``vortex-like'' excitations carrying charge (i.e.~one with a low energy pole in the conductivity and one with a low energy zero) \cite{Herzog:2007ij,Myers:2010pk}, and we expect an analogous duality to hold for our neutral theory close to the self-dual point $m=\sqrt{8}\pi T$. Near the self-dual point (where the thermodynamic energy vanishes), our system has either a small negative or positive thermodynamic energy. To get a better understanding of the physics of this neutral state, it would be very interesting to find a (non-holographic) field theory with heat conduction properties of this type, along the lines of \cite{Herzog:2007ij}. Moreover, at the self-dual point, an extra SL(2,$\mathbb{R}$)$\times$SL(2,$\mathbb{R}$) is present and allows to solve exactly for the Green's functions for all $\omega$ and $k$. Assuming it is still given by the usual Kubo formula, a `shear viscosity' can be calculated exactly $\eta\approx0.325\times s/4\pi$, and violates the KSS bound. Such violations are due to translational symmetry breaking and will also occur in states with different values of $m$.

On a related note, the similarities between our holographic model and massive gravity (where the parameter $m$ now becomes related to the graviton mass) have been discussed in \cite{Andrade:2013gsa,Taylor:2014tka}. It is known that linearized massive gravity in de Sitter backgrounds becomes partially massless at a specific value of the graviton mass, with a corresponding emergent self-duality \cite{Deser:2013xb,Hinterbichler:2014xga}. It would be interesting to understand better the links between the two theories, and possibly the emergence of self-duality in massive gravity.

Finally, in light of recent memory matrix studies of transport properties, we highlighted some interesting properties of probe brane theories and neutral theories with higher-derivative couplings between the gauge field and the Weyl tensor. In particular, probe brane theories at low temperatures typically exhibit coherent peaks in their low energy transport, as can the higher-derivative theories (although in limits where effective field theory breaks down or causality is violated). According to memory matrix calculations, an almost-conserved quantity overlapping with the charge current will produce a coherent peak in the low energy electrical conductivity \cite{forster1990hydrodynamic, Hartnoll:2007ih, Hartnoll:2012rj, Mahajan:2013cja}. But is the converse true? Does the existence of coherent peaks, e.g.~in probe brane systems, indicate an approximate conservation law at work? If so, this conservation law should also explain the nature of the holographic zero sound excitation in these systems.

\begin{acknowledgments}
We are grateful to Sean Hartnoll, Diego Hofman, Elias Kiritsis, Andrei Parnachev, Anastasios Petkou, Koenraad Schalm, David Tong and Jan Zaanen for helpful discussions. We also thank Sean Hartnoll, Chris Herzog and Andy Lucas for comments on the manuscript. The work of R.D. is supported by a VIDI grant from NWO, the Netherlands Organisation for Scientific Research. The work of B.G. is supported by the Marie Curie International Outgoing
Fellowship nr 624054 within the 7$^\textrm{th}$ European Community Framework Programme FP7/2007-2013.
\end{acknowledgments}

\clearpage

\appendix

\section{Details of numerical calculations}
\label{sec:numericalappendix}

The linear response properties of quantum field theory states are encoded in their two-point retarded Green's functions. To calculate these from the gravitational dual, we must solve the equations of motion for the linear perturbations around the corresponding black brane solution. The two-point retarded Green's function of an operator $\mathcal{O}$, dual to a bulk field $\phi_\mathcal{O}$, is then given by the value of the on-shell gravitational action at quadratic order in perturbations, after imposing the boundary conditions that the `source' term of each field is zero, except the source term of $\phi_{\mathcal{O}}$ which should be equal to one, and that all field perturbations are ingoing at the black brane horizon. We refer the reader to \cite{Son:2002sd,Kaminski:2009dh} for more technical details. 

In the following subsection we will give the relevant equations of motion for the fluctuations of each field, which are typically coupled to one another. After determining the appropriate decoupled field perturbations, we solve their equations of motion by numerical integration from the black brane horizon to the spacetime boundary and use these solutions to compute the retarded Green's functions, and their poles, that we present in the main body of the paper.

\subsection{Neutral axionic theory}

In the neutral axionic theory with action (\ref{eq:axiontheoryaction}), we are primarily concerned with the linear perturbations of $\delta g_{t{x^i}}$, which are dual to the momentum density operators $T^{t{x^i}}$ of the dual field theory. In general, we are interested in these perturbations at non-zero frequency $\omega$ and wavenumber $\vec{k}$. We choose our co-ordinates such that the wavenumber $\vec{k}$ points in the $x$-direction. The linear perturbations of the metric $\delta g_{\mu\nu}\equiv h_{\mu\nu}$ and of the scalar fields $\delta\phi_i$ form two decoupled sets: those related to longitudinal transport ($h_{tt}$, $h_{xx}$, $h_{yy}$, $h_{rr}$, $h_{xt}$, $h_{xr}$, $h_{tr}$, $\delta\phi_1$), and those related to transverse transport ($h_{yt}$, $h_{yx}$, $h_{yr}$, $\delta\phi_2$). The longitudinal fields obey the coupled equations of motion
\begin{align}
0=\;&\frac{d}{dr}\left[r^4f\left(\delta\phi_1'-mh^x_r\right)\right]+\frac{1}{f}\left(\omega^2-k^2f\right)\delta\phi_1-\frac{i\omega m}{f}h^x_t+\frac{ikm}{2}\left(h^t_t-h^x_x+h^y_y+h^r_r\right),\\
0=\;&\frac{d}{dr}\left[r^4f\left({h^x_x}'-{h^y_y}'-2ikh^x_r\right)\right]+2ikm\delta\phi_1+\frac{1}{f}\left(\omega^2-m^2f\right)\left(h^x_x-h^y_y\right)+\frac{2\omega k}{f}h^x_t\\
&-k^2\left(h^t_t+h^r_r\right),\nonumber\\
0=\;&\frac{d}{dr}\left[r^4\left({h^x_t}'+i\omega h^x_r\right)\right]-\frac{ik}{f}{h^r_t}'+\frac{\omega k}{f}\left(h^y_y+h^r_r\right)-\frac{i\omega m}{f}\delta\phi_1-\frac{m^2}{f}h^x_t,\\
0=\;&i\omega\left({h^x_x}'+{h^y_y}'\right)+ik{h^x_t}'+\omega kh^x_r+\frac{k^2}{r^4f}h^r_t-\frac{i\omega f'}{2f}\left(h^x_x+h^y_y\right)-\frac{2i\omega}{r}h^r_r-ik\frac{f'}{f}h^x_t,\label{eq:axionconstraint1}\\
0=\;&ik\left({h^t_t}'+{h^y_y}'\right)-\frac{i\omega}{f}{h^x_t}'+\frac{\omega^2}{f}h^x_r+\frac{\omega k}{r^4f^2}h^r_t+\frac{ikf'}{2f}h^t_t-ik\left(\frac{f'}{2f}+\frac{2}{r}\right)h^r_r-m^2h^x_r+m\delta\phi_1',\label{eq:axionconstraint2}\\
0=\;&2r^3f^2{h^t_t}'+2r^4f^2\left(\frac{f'}{4f}+\frac{1}{r}\right)\left({h^x_x}'+{h^y_y}'\right)+ikmf\delta\phi_1-2ikr^4f^2\left(\frac{f'}{2f}+\frac{2}{r}\right)h^x_r -\frac{4i\omega}{r}h^r_t\label{eq:axionconstraint3}\\
&+\omega^2\left(h^x_x+h^y_y\right)-k^2f\left(h^t_t+h^y_y\right)+2\omega kh^x_t-\frac{1}{2}m^2f\left(h^x_x+h^y_y\right)+f\left(m^2-6r^2\right)h^r_r,\nonumber\\
0=\;&r^4f^2\left({h^x_x}''+{h^y_y}''\right)+\frac{1}{2}r^3f\left(rf'+8f\right)\left({h^x_x}'+{h^y_y}'\right)-2ikr^4f^2{h^x_r}'-2r^3f^2{h^r_r}'\\
&-ikr^3f\left(rf'+8f\right)h^x_r+ikmf\delta\phi_1-k^2f\left(h^y_y+h^r_r\right)-\frac{1}{2}m^2f\left(h^x_x+h^y_y\right)\nonumber\\
&+\frac{f}{3}\left[3m^2-18r^2+6r^2\left(3f+rf'\right)\right]h^t_t-\frac{r^2f}{3}\left(6rf'+18f\right)h^r_r,\nonumber\\
0=\;&2r^4f^2{h^t_t}''+r^3f\left(3rf'+8f\right){h^t_t}'+r^4f^2\left({h^x_x}''+{h^y_y}''\right)-2ikr^4f^2{h^x_r}'-4i\omega{h^r_t}'\\
&+r^3f\left(rf'+4f\right)\left({h^x_x}'+{h^y_y}'-{h^r_r}'\right)-2ikr^3f\left(rf'+4f\right)h^x_r+2i\omega\frac{f'}{f}h^r_t+2\omega^2 h^r_r\nonumber\\
&+\omega^2\left(h^x_x+h^y_y\right)-k^2f\left(h^t_t+h^r_r\right)+2\omega kh^x_t-\frac{2r^2f}{3}\left(18rf'+18f+3r^2f''\right)h^r_r,\nonumber
\end{align}
and the transverse fields obey the coupled equations of motion
\begin{align}
0=\;&\frac{d}{dr}\left[r^4f\left({h^y_x}'-ikh^y_r\right)\right]+\frac{\omega}{f}\left(\omega h^y_x+kh^y_t\right)-m^2h^y_x+ikm\delta\phi_2,\\
0=\;&\frac{d}{dr}\left[r^4\left({h^y_t}'+i\omega h^y_r\right)\right]-\frac{k}{f}\left(\omega h^y_x+kh^y_t\right)-\frac{m^2}{f}h^y_t-\frac{i\omega m}{f}\delta\phi_2,\\
0=\;&\frac{d}{dr}\left[r^4f\left(\delta\phi_2'-mh^y_r\right)\right]+\frac{1}{f}\left(\omega^2-k^2f\right)\delta\phi_2-\frac{m}{f}\left(i\omega h^y_t+ikfh^y_x\right),\\
0=\;&i\omega{h^y_t}'+ikf{h^y_x}'-\left(\omega^2-m^2f-k^2f\right)h^y_r-mf\delta\phi_2', \label{eq:axionconstraint4}
\end{align}
where we raise indices using the background metric (\ref{eq:axiontheorymetricsolution}) and primes denote derivatives with respect to $r$. Longitudinal and transverse transport in this theory are therefore, in general, independent of one another.

Despite the large number of fields and equations, there are only four independent degrees of freedom in the system: two in the longitudinal sector and two in the transverse sector. The multitude of field equations are a redundancy due to the diffeomorphism symmetry of the theory, that can be removed by working directly with combinations of the fields that are invariant under linearised diffeomorphisms $h_{\mu\nu}\rightarrow h_{\mu\nu}+\nabla_\mu\xi_\nu+\nabla_\nu\xi_\mu$ and $\delta\phi_i\rightarrow\delta\phi_i+m\xi^i$. By choosing these variables carefully, we can identify a decoupled, gauge-invariant field perturbation in each sector that determines the two-point function of the appropriate component of the momentum density $P_i=T^{tx^{i}}$. This decoupling is extremely advantageous when it comes to solving these equations. In the longitudinal sector, the field
\begin{align}
\label{eq:axionlongitudinalgivariable}
\psi_\parallel\equiv&\frac{r}{m^2+k^2}\left[m\left(h^x_x-h^y_y\right)-2ik\delta\phi_1\right]\\
&-\frac{2r^4f}{m\left(k^2+r^3f'\right)}\Bigl[{h^x_x}'+{h^y_y}'-\frac{2}{r}h^r_r-2ikh^x_r-\frac{k^2+r^3f'}{r^3f}h^y_y\Bigr]\nonumber,
\end{align}
obeys the equation of motion
\begin{equation}
\label{eq:axionlongitudinalgiequation}
\frac{d}{dr}\left[r^2f\psi_\parallel'\right]+\left(\frac{\omega^2-k^2f}{r^2f}+V(r)\right)\psi_\parallel=0,
\end{equation}
where
\begin{align}
V(r)=-\frac{3r_0\left(m^2-2r_0^2\right)}{2r^3\left[2k^2r+m^2\left(2r-3r_0\right)+6r_0^3\right]^2}\Bigl[&4k^4r^2+m^4\left(-4r^2+6rr_0-3r_0^2\right)\\
&+12m^2r_0\left(r^3-rr_0^2+r_0^3\right)-12r_0^3\left(2r^3+r_0^3\right)\Bigr]\nonumber,
\end{align}
and controls the two point function of $T^{xt}$. In the transverse sector, the field
\begin{equation}
\label{eq:axiontransversegivariable}
\psi_\perp\equiv r^3\left({h^y_t}'+i\omega h^y_r\right),
\end{equation}
obeys the equation of motion
\begin{equation}
\label{eq:axiontransversegiequation}
\frac{d}{dr}\left[r^2f\psi_\perp'\right]+\frac{\omega^2-k^2f-m^2f+r^3ff'}{r^2f}\psi_\perp=0,
\end{equation}
and controls the two point function of $T^{yt}$. We note that it is also possible to write down decoupled equations of motion for the other two independent degrees of freedom of the system, but for conciseness we will not present them here.

To compute the Green's functions, we must determine the on-shell gravitational action at quadratic order in the fluctuations, after including the Gibbons-Hawking boundary term, and boundary counterterms
\begin{equation}
\label{eq:axionandmetriccounterterms}
S_{GH}+S_{ct}=\int d^3x\sqrt{-\gamma}\left(-2K-4-\mathcal{R}[\gamma]+\sum_{i=1}^{2}\frac{1}{2}\partial_\mu\phi_i\partial^\mu\phi_i\right),
\end{equation}
where $\gamma$ is the induced metric and the integral is over the boundary of the spacetime. On-shell, we can replace the fields $h^r_t$, $h^r_r$, $h^x_r$ and $h^y_r$ by solving the constraint equations (\ref{eq:axionconstraint1}), (\ref{eq:axionconstraint2}), (\ref{eq:axionconstraint3}) and (\ref{eq:axionconstraint4}) for these fields. Furthermore, as the on-shell action is a boundary term, we can replace all of the remaining fields by their near-boundary expansions
\begin{equation}
\label{eq:nearboundartperturbationexpansion}
h^x_t={h^x_t}^{(0)}+\frac{{h^x_t}^{(1)}}{r}+\frac{{h^x_t}^{(2)}}{r^2}+\frac{{h^x_t}^{(3)}}{r^3}+\frac{{h^x_t}^{(4)}}{r^4}+\ldots,
\end{equation}
and similarly for the other metric and scalar field fluctuations. On-shell, only two of these boundary coefficients are independent: the `source' term ${h^x_t}^{(0)}$ and the `expectation value' term ${h^x_t}^{(3)}$. The result of this is that the on-shell action can be written as a sum of terms which are quadratic in the source and expectation value coefficients. The expectation value terms should be regarded as functions of the source terms, with the precise relationship between them being fixed by demanding ingoing boundary conditions at the horizon. This relationship will be determined via numerical integration.

We have computed the quadratic on-shell action in this form in full generality but, to compute the retarded Green's functions of $T^{tx}$ and $T^{ty}$, we only need to know the on-shell action with all source terms vanishing, except ${h^x_t}^{(0)}$ and ${h^y_t}^{(0)}$. This is
\begin{align}
S=&\int\frac{d\omega dk}{4\pi^2}\Biggl(\frac{1}{4k^4-4m^2\omega^2+4\omega^4+k^2\left(3m^2-8\omega^2\right)}\Bigl[\omega {h^x_t}^{(0)}\Bigl\{-3k\left(k^2+m^2\right){h^t_t}^{(3)}\nonumber\\
&+\frac{3}{2}k\left(2k^2+m^2-2\omega^2\right){h^y_y}^{(3)}+\frac{3}{2}k\left(m^2+2\omega^2\right){h^x_x}^{(3)}+6\omega\left(k^2+m^2\right){h^x_t}^{(3)}\nonumber\\
&-3im\left(k^2-2\omega^2\right)\delta\phi_1^{(3)}\Bigr\}-\frac{1}{4}r_0\left(m^2-2r_0^2\right)\Bigl\{8\omega^2\left(m^2-\omega^2\right)+k^2\left(3m^2+10\omega^2\right)\nonumber\\
&+4k^4\Bigr\}{h^x_t}^{(0)}{h^x_t}^{(0)}\Bigr]-\frac{3}{2\left(k^2+m^2-\omega^2\right)}{h^y_t}^{(0)}\Bigl\{\left(k^2+m^2\right){h^y_t}^{(3)}+\omega k{h^y_x}^{(3)}+im\omega\delta\phi_2^{(3)}\Bigr\}\nonumber\\
&+\frac{1}{2}r_0\left(m^2-2r_0^2\right){h^y_t}^{(0)}{h^y_t}^{(0)}\Biggr),
\end{align}
where, in each quadratic pair of field perturbations, the first has argument $(-\omega,-k)$ and the second has argument $(\omega,k)$. In general, we will neglect the terms ~${h^x_t}^{(0)}(-\omega,-k){h^x_t}^{(0)}(\omega,k)$ and ~${h^y_t}^{(0)}(-\omega,-k){h^y_t}^{(0)}(\omega,k)$ because these are real contact terms, whose only effect is to shift the real part of the dual Green's function by their prefactor. In particular, these terms do not affect the real (dissipative) parts of the conductivities.

The final step required to determine the Green's functions is to evaluate the specific combinations of expectation value terms, on a solution that is ingoing at the horizon and which has only one source term (${h^x_t}^{(0)}=1$ or ${h^y_t}^{(0)}=1$) non-vanishing at the boundary. This is where the advantage of the gauge-invariant variables (\ref{eq:axionlongitudinalgivariable}) and (\ref{eq:axiontransversegivariable}) is manifest. Near the boundary, these have the expansions
\begin{align}
\psi_\perp\left(r\rightarrow\infty\right)\rightarrow&\left(k^2+m^2\right){h^y_t}^{(0)}+\omega k{h^y_x}^{(0)}+im\omega\delta\phi_2^{(0)}\\
&-\frac{3}{\left(k^2+m^2-\omega^2\right)r}\left[\left(k^2+m^2\right){h^y_t}^{(3)}+\omega k{h^y_x}^{(3)}+im\omega\delta\phi_2^{(3)}\right]+O(1/r^2)\nonumber,
\end{align}
and
\begin{align}
\label{eq:axionlongitudinalgibdyexpansion}
\psi_\parallel\left(r\rightarrow\infty\right)\rightarrow&-\frac{8k}{m\left(k^2+m^2\right)}\frac{1}{4k^4-4m^2\omega^2+4\omega^4+k^2\left(3m^2-8\omega^2\right)}\times\\
&\Bigl\{\frac{3}{2}k\left(k^2+m^2\right){h^t_t}^{(3)}-\frac{3}{4}k\left(2k^2+m^2-2\omega^2\right){h^y_y}^{(3)}-\frac{3}{4}k\left(m^2+2\omega^2\right){h^x_x}^{(3)}\nonumber\\
&-3\omega\left(k^2+m^2\right){h^x_t}^{(3)}+\frac{3}{2}im\left(k^2-2\omega^2\right)\delta\phi_1^{(3)}\nonumber\\
&-\frac{3}{8}r_0\left(m^2-2r_0^2\right)\left(3k^2+2m^2-2\omega^2\right)\left(k{h^t_t}^{(0)}-2\omega{h^x_t}^{(0)}\right)\Bigr\}+O(1/r),\nonumber
\end{align}
respectively. Firstly, we will deal with the transverse correlator $G^R_{T^{ty}T^{ty}}$. A solution in which ${h^y_t}^{(0)}=1$ and all other source terms vanish, is a solution in which $\psi_\perp(r\rightarrow\infty)\rightarrow k^2+m^2$. It is straightforward to find such a solution: one just integrates the equation of motion (\ref{eq:axiontransversegiequation}) from the horizon, with ingoing boundary conditions, to the boundary. Multiplying this solution by $\left(k^2+m^2\right)$ and dividing by its value at the boundary will yield a solution with precisely the boundary conditions required, and thus
\begin{equation}
\left(k^2+m^2\right)\frac{r^2\psi_\perp'(r)}{\psi_\perp(r)}\Biggr|_{r\rightarrow\infty}=\frac{3}{k^2+m^2-\omega^2}\left[\left(k^2+m^2\right){h^y_t}^{(3)}+\omega k{h^y_x}^{(3)}+im\omega\delta\phi_2^{(3)}\right],
\end{equation} 
where the right hand side is evaluated with the boundary conditions that all fields are ingoing at the horizon and that the only non-vanishing source term at the boundary is ${h^y_t}^{(0)}=1$. Thus, up to contact terms,
\begin{equation}
\label{eq:axiontransversecorrelatoreqn}
G^R_{T^{ty}T^{ty}}(\omega,k)=\left(k^2+m^2\right)\frac{r^2\psi_\perp'(r)}{\psi_\perp(r)}\Biggr|_{r\rightarrow\infty}.
\end{equation}

A similar procedure works for the longitudinal retarded Green's functions. Near the boundary, the gauge-invariant variable (\ref{eq:axionlongitudinalgivariable}) has the expansion (\ref{eq:axionlongitudinalgibdyexpansion}). Since the leading order term involves the expectation values (${h^x_t}^{(3)}$ etc.), there is no simple expression for the boundary value of $\psi_\parallel$ in a solution in which all source terms vanish except ${h^x_t}^{(0)}$. However, one can define a new gauge-invariant longitudinal field
\begin{equation}
\psi_\parallel^{\text{new}}\equiv r^2f\psi_\parallel'(r)+\frac{3r_0\left(m^2-2r_0^2\right)}{2\left(k^2+m^2\right)}\psi_\parallel(r),
\end{equation}
whose near boundary expansion is
\begin{equation}
\psi_\parallel^{\text{new}}\left(r\rightarrow\infty\right)\rightarrow\frac{2\omega k}{m}{h^x_t}^{(0)}+O(1/r),
\end{equation}
when all other source terms vanish. Using similar arguments to before, one can then deduce that up to contact terms,
\begin{equation}
\label{eq:axionlongitudinalcorrelatoreqn}
G^R_{T^{tx}T^{tx}}(\omega,k)=\omega^2\left(k^2+m^2\right)\frac{\psi_\parallel(r)}{r^2f\psi_\parallel'(r)+\frac{3r_0\left(m^2-2r_0^2\right)}{2\left(k^2+m^2\right)}\psi_\parallel(r)}\Biggr|_{r\rightarrow\infty}.
\end{equation}
Using this approach, it is straightforward to show (although we omit the details) that the gauge-invariant field (\ref{eq:axionlongitudinalgivariable}) also controls the retarded Green's function of the energy density $T^{tt}$, and that it automatically obeys the Ward identity $G^R_{T^{tt}T^{tt}}=k^2G^R_{T^{tx}T^{tx}}/\omega^2$, due to this gauge-invariant formulation.

To find the poles of the conductivities as shown in the main text, we simply need to find the poles of the Green's functions i.e.~where the denominators of the right hand sides of (\ref{eq:axiontransversecorrelatoreqn}) and (\ref{eq:axionlongitudinalcorrelatoreqn}) vanish. To compute the real (dissipative) parts of the conductivities as shown in the main text, we simply take the imaginary parts of the Green's functions (\ref{eq:axiontransversecorrelatoreqn}) and (\ref{eq:axionlongitudinalcorrelatoreqn}) and divide by $\omega$ (as per the definitions (\ref{eq:hydroheatconductivity}) and (\ref{eq:transversemomentumconductivitydefninhydro})). In all cases, we have shown, the imaginary part of the subtracted terms in (\ref{eq:hydroheatconductivity}) and (\ref{eq:transversemomentumconductivitydefninhydro}) vanish, and so we do not need to explicitly do this subtraction. To determine the spatially uniform AC heat conductivity $\kappa\left(\omega,k=0\right)$, one can easily take the $k=0$ limits of the transverse equations (\ref{eq:axiontransversegiequation}) and (\ref{eq:axiontransversecorrelatoreqn}).

\section{Calculation of Green's functions at self-dual point}
\label{sec:selfdualappendix}

When $m=\sqrt{2}\,r_0$, the gravitational theory exhibits an enhanced symmetry that allows us to explicitly solve the perturbation equations. In the following subsections, we outline how the solutions to these equations at the self-dual point $m=\sqrt{2}\,r_0$ are related to the two-point retarded Green's functions in the dual field theory.

\subsection{Ward identities}

There are 8 independent operators in the sector of the field theory state which is dual to the metric and scalar field perturbations of the gravitational theory. These operators can be divided into transverse and longitudinal sectors ($\left\{P^{y}\equiv T^{ty},T^{xy},\Phi_2\right\}$ and $\left\{P^x\equiv T^{xt},T^{tt},T^{xx},T^{yy},\Phi_1\right\}$ respectively), which decouple from one another i.e.~cross correlators between an operator in the transverse sector and an operator in the longitudinal sector vanish. Within each sector, there is a further decoupling such that
\begin{equation}
G^R_{P^x\Phi_1}\left(\omega,k\right)=G^R_{\Phi_1P^x}\left(\omega,k\right)=0,\;\;\;\;\;\;\;\text{and}\;\;\;\;\;\;\; G^R_{P^y\Phi_2}\left(\omega,k\right)=G^R_{\Phi_2P^y}\left(\omega,k\right)=0.
\end{equation}
This is dual to the decoupling of the corresponding perturbations in the gravitational theory. It would be worthwhile to find a good field theoretical reason for this decoupling.

Even after imposing these constraints, there are naively many more two-point Green's functions than the number of gravitational degrees of freedom (four). This mismatch can be easily resolved: there are really only four independent Green's functions, and the others are related to them by Ward identities arising from the relations \cite{Taylor:2014tka,Andrade:2013gsa}
\begin{equation}
\partial_\mu T^{\mu\nu}=-\sum_{i=1}^{2}\mathcal{O}_i\partial^\nu\mathcal{J}_i,\;\;\;\;\;\;\;\;\;\;\;\;\;\;\;\;\;\;\; T^{\mu}_{\mu}=0,
\end{equation}
where $\mathcal{J}_i$ is the source for the operator $\mathcal{O}_i$. Using these relations, we can express all of the Green's functions in terms of just $G^R_{P^xP^x}$, $G^R_{P^yP^y}$, $G^R_{\Phi_1\Phi_1}$ and $G^R_{\Phi_2\Phi_2}$ as follows (up to contact terms)
\begin{align}
\label{eq:wardidentitieslist}
G^R_{P^yT^{xy}}\left(\omega,k\right)&=\frac{\omega k}{\left(k^2+2r_0^2\right)}G^R_{P^yP^y}\left(\omega,k\right),\;\;\;\;\;\;\; G^R_{T^{xy}\Phi_2}\left(\omega,k\right)=\frac{\sqrt{2}\,r_0}{\left(k^2+2r_0^2\right)}G^R_{\Phi_2\Phi_2}\left(\omega,k\right),\\
G^R_{T^{xy}T^{xy}}\left(\omega,k\right)&=\frac{\omega^2k^2}{\left(k^2+2r_0^2\right)^2}G^R_{P^yP^y}\left(\omega,k\right)+\frac{2r_0^2}{\left(k^2+2r_0^2\right)^2}G^R_{\Phi_2\Phi_2}\left(\omega,k\right),\nonumber\\
G^R_{P^xT^{xx}}\left(\omega,k\right)&=\frac{k\left(r_0^2+\omega^2\right)}{\omega\left(k^2+2r_0^2\right)}G^R_{P^xP^x}\left(\omega,k\right),\;\;\;\;\;\;\; G^R_{P^xT^{yy}}\left(\omega,k\right)=\frac{k\left(k^2+r_0^2-\omega^2\right)}{\omega\left(k^2+2r_0^2\right)}G^R_{P^xP^x}\left(\omega,k\right),\nonumber\\
G^R_{T^{xx}T^{xx}}\left(\omega,k\right)&=\frac{r_0^2}{2\left(k^2+2r_0^2\right)^2}G^R_{\Phi_1\Phi_1}\left(\omega,k\right)+\frac{k^2\left(r_0^2+\omega^2\right)^2}{\omega^2\left(k^2+2r_0^2\right)^2}G^R_{P^xP^x}\left(\omega,k\right),\nonumber\\
G^R_{T^{yy}T^{yy}}\left(\omega,k\right)&=\frac{r_0^2}{2\left(k^2+2r_0^2\right)^2}G^R_{\Phi_1\Phi_1}\left(\omega,k\right)+\frac{k^2\left(k^2+r_0^2-\omega^2\right)^2}{\omega^2\left(k^2+2r_0^2\right)^2}G^R_{P^xP^x}\left(\omega,k\right),\nonumber\\
G^R_{T^{xx}T^{yy}}\left(\omega,k\right)&=-\frac{r_0^2}{2\left(k^2+2r_0^2\right)^2}G^R_{\Phi_1\Phi_1}\left(\omega,k\right)+\frac{k^2\left(r_0^2+\omega^2\right)\left(k^2+r_0^2-\omega^2\right)}{\omega^2\left(k^2+2r_0^2\right)^2}G^R_{P^xP^x}\left(\omega,k\right),\nonumber\\
G^R_{T^{xx}\Phi_1}\left(\omega,k\right)&=\frac{r_0}{\sqrt{2}\left(k^2+2r_0^2\right)}G^R_{\Phi_1\Phi_1}\left(\omega,k\right),\;\;\;\;\;\;\; G^R_{T^{yy}\Phi_1}\left(\omega,k\right)=-\frac{r_0}{\sqrt{2}\left(k^2+2r_0^2\right)}G^R_{\Phi_1\Phi_1}\left(\omega,k\right),\nonumber
\end{align}
and $G^R_{T^{tt}\mathcal{O}}\left(\omega,k\right)=kG^R_{P^x\mathcal{O}}\left(\omega,k\right)/\omega$, due to energy conservation.

\subsection{On-shell action and Green's functions}

As outlined in the main text, all of the gravitational equations can be reduced to just two different equations ((\ref{eq:firstmetricselfdualeqn}) and (\ref{eq:secondmetricselfdualeqn})) at the self-dual point. We denote the solutions to these equations as $\varphi$ and $\Psi$ respectively. After solving them, and imposing ingoing boundary conditions at the horizon, their behaviour near the boundary can be used to determine the Green's functions of the dual field theory. This is because their boundary behaviour gives the boundary behaviour of the fluctuations of the metric and scalar fields, and it is these which control the aforementioned Green's functions. 

Specifically, near the boundary of the spacetime, the fields $\varphi_{p_x}$ and $\varphi_{p_y}$, defined in equations (\ref{eq:varphipxdefn}) and (\ref{eq:varphipydefn}), are
\begin{align}
\label{eq:selfdualnearboundaryexpansionsofGIfields1}
\varphi_{p_x}=&\frac{3\sqrt{2}k}{r_0\left[2k^4-4r_0^2\omega^2+2\omega^4+k^2\left(3r_0^2-4\omega^2\right)\right]}\Biggl\{k\left(\omega^2+r_0^2\right){h^x_x}^{(3)}-k\left(k^2+2r_0^2\right){h^t_t}^{(3)}\\
&-i\sqrt{2}k^2r_0\delta\phi_1^{(3)}+2\omega\left(k^2+2r_0^2\right){h^x_t}^{(3)}+i2\sqrt{2}\,r_0\omega^2\delta\phi_1^{(3)}+k\left(k^2+r_0^2-\omega^2\right){h^y_y}^{(3)}\Biggr\},\nonumber\\
&+\frac{k}{2r_0r}\Biggl\{-\sqrt{2}kr_0^2{h^x_x}^{(0)}+\sqrt{2}k\left(k^2+2r_0^2\right){h^t_t}^{(0)}+2ik^2r_0\delta\phi_1^{(0)}-2\sqrt{2}k^2\omega{h^x_t}^{(0)}\nonumber\\
&-4\sqrt{2}\,r_0^2\omega{h^x_t}^{(0)}-\sqrt{2}k\omega^2{h^x_x}^{(0)}-4ir_0\omega^2\delta\phi_1^{(0)}-\sqrt{2}k\left(k^2+r_0^2-\omega^2\right){h^y_y}^{(0)}\Biggr\}+O(1/r^2),\nonumber\\
\varphi_{p_y}=&\Biggl\{\left(k^2+2r_0^2\right){h^y_t}^{(0)}+\omega k{h^y_x}^{(0)}+i\sqrt{2}\,r_0\omega\delta\phi_2^{(0)}\Biggr\}\\
&-\frac{3}{\left(k^2+2r_0^2-\omega^2\right)r}\Biggl\{\left(k^2+2r_0^2\right){h^y_t}^{(3)}+\omega k{h^y_x}^{(3)}+i\sqrt{2}\,r_0\omega\delta\phi_2^{(3)}\Biggr\}+O(1/r^2),\nonumber
\end{align}
where ${h^x_t}^{(i)}$ etc. are the coefficients in the near boundary expansions of the field perturbations (\ref{eq:nearboundartperturbationexpansion}). These fields both obey the $\varphi$ equation of motion (\ref{eq:firstmetricselfdualeqn}). Similarly, the fields $\Psi_x$ and $\Psi_y$, defined in equations (\ref{eq:psixdefn}) and (\ref{eq:psiydefn}), have near-boundary expansions
\begin{align}
\label{eq:selfdualnearboundaryexpansionsofGIfields2}
\Psi_x=&\left\{\sqrt{2}\,r_0\left({h^x_x}^{(0)}-{h^y_y}^{(0)}\right)-2ik\delta\phi_1^{(0)}\right\}\left(1-\frac{\left(k^2+2r_0^2-\omega^2\right)}{2r^2}\right)\\
&+\frac{1}{r^3}\left\{\sqrt{2}\,r_0\left({h^x_x}^{(3)}-{h^y_y}^{(3)}\right)-2ik\delta\phi_1^{(3)}\right\}+O(1/r^4),\nonumber\\
\Psi_y=&\left\{\sqrt{2}\,r_0{h^y_x}^{(0)}-ik\delta\phi_2^{(0)}\right\}\left(1-\frac{k^2+2r_0^2-\omega^2}{2r^2}\right)+\frac{1}{r^3}\left\{\sqrt{2}\,r_0{h^y_x}^{(3)}-ik\delta\phi_2^{(3)}\right\}\\
&+O(1/r^4).\nonumber
\end{align}
These fields both obey the $\Psi$ equation of motion (\ref{eq:secondmetricselfdualeqn}).

To determine the precise relation between the boundary values of the fields and the retarded Green's functions, we evaluate the total action (given in equations (\ref{eq:axiontheoryaction}) and (\ref{eq:axionandmetriccounterterms})) on-shell at the self-dual point $m=\sqrt{2}\,r_0$. Working to quadratic order in the perturbations, it is
\begin{align}
S=&\int\frac{d\omega dk}{4\pi^2}\Biggl\{\frac{1}{\left[4k^4-8r_0^2\omega^2+4\omega^4+2k^2\left(3r_0^2-4\omega^2\right)\right]}\Biggl[{h^x_x}^{(0)}\Biggl\{-\frac{3}{2}k^2\left(r_0^2+\omega^2\right){h^t_t}^{(3)}\\
&-\frac{3}{4}\left\{2k^2\left(r_0^2-\omega^2\right)-2\omega^2\left(2r_0^2-\omega^2\right)\right\}{h^y_y}^{(3)}+\frac{3}{2}\left(2r_0^2k^2-2r_0^2\omega^2+\omega^4\right){h^x_x}^{(3)}\nonumber\\
&+3k\omega\left(r_0^2+\omega^2\right){h^x_t}^{(3)}-\frac{3}{\sqrt{2}}ikr_0\left(2k^2-3\omega^2\right)\delta\phi_1^{(3)}\Biggr\}\nonumber\\
&+{h^y_y}^{(0)}\Biggl\{-\frac{3}{2}k^2\left(k^2+r_0^2-\omega^2\right){h^t_t}^{(3)}+\frac{3}{2}\left(k^2-\omega^2\right)\left(k^2+2r_0^2-\omega^2\right){h^y_y}^{(3)}\nonumber\\
&-\frac{3}{2}\left\{k^2\left(r_0^2-\omega^2\right)-\omega^2\left(2r_0^2-\omega^2\right)\right\}{h^x_x}^{(3)}+3k\omega\left(k^2+r_0^2-\omega^2\right){h^x_t}^{(3)}+\frac{3}{\sqrt{2}}ikr_0\left(k^2-\omega^2\right)\delta\phi_1^{(3)}\Biggr\}\nonumber\\
&+\left({h^t_t}^{(0)}-\frac{2\omega}{k}{h^x_t}^{(0)}\right)\Biggl\{\frac{3}{2}k^2\left(k^2+2r_0^2\right){h^t_t}^{(3)}-\frac{3}{2}k^2\left(k^2+r_0^2-\omega^2\right){h^y_y}^{(3)}-\frac{3}{2}k^2\left(\omega^2+r_0^2\right){h^x_x}^{(3)}\nonumber\\
&-3k\omega\left(k^2+2r_0^2\right){h^x_t}^{(3)}+\frac{3}{\sqrt{2}}ikr_0\left(k^2-2\omega^2\right)\delta\phi_1^{(3)}\Biggr\}\nonumber\\
&+\delta\phi_1^{(0)}\Biggl\{-\frac{3}{\sqrt{2}}ikr_0\left(k^2-2\omega^2\right){h^t_t}^{(3)}-\frac{3}{\sqrt{2}}ikr_0\left(k^2-\omega^2\right){h^y_y}^{(3)}+\frac{3}{\sqrt{2}}ikr_0\left(2k^2-3\omega^2\right){h^x_x}^{(3)}\nonumber\\
&+3\sqrt{2}ir_0\omega\left(k^2-2\omega^2\right){h^x_t}^{(3)}+6\left(k^2-\omega^2\right)^2\delta\phi_1^{(3)}\Biggr\}\Biggr]\nonumber\\
&+\frac{3}{2\left(k^2+2r_0^2-\omega^2\right)}\Biggl[{h^y_x}^{(0)}\left\{\left(2r_0^2-\omega^2\right){h^y_x}^{(3)}-k\omega{h^y_t}^{(3)}-i\sqrt{2}\,r_0k\delta\phi_2^{(3)}\right\}\nonumber\\
&-{h^y_t}^{(0)}\left\{\left(k^2+2r_0^2\right){h^y_t}^{(3)}+k\omega{h^y_x}^{(3)}+i\sqrt{2}\,r_0\omega\delta\phi_2^{(3)}\right\}\nonumber\\
&+\delta\phi_2^{(0)}\left\{i\sqrt{2}\,r_0\omega{h^y_t}^{(3)}+i\sqrt{2}\,r_0k{h^y_x}^{(3)}+\left(k^2-\omega^2\right)\delta\phi_2^{(3)}\right\}\Biggr]\Biggr\},\nonumber
\end{align}
where, in each quadratic pair of field perturbations, the first has argument $(-\omega,-k)$ and the second has argument $(\omega,k)$. With this action, the usual holographic procedure (as described in appendix \ref{sec:numericalappendix}) can be used to find expressions for each two-point Green's function in terms of the near-boundary coefficients of the fields (${h^x_t}^{(3)}$ etc.). These coefficients can be related to the near-boundary coefficients of the fields $\varphi=\sum_i\varphi^{(i)}r^{-i}$ and $\Psi=\sum_i\Psi^{(i)}r^{-i}$ using equations (\ref{eq:selfdualnearboundaryexpansionsofGIfields1}) and (\ref{eq:selfdualnearboundaryexpansionsofGIfields2}), yielding (for example) the results (\ref{eq:selfdualcorrelator1intermsofbuildingblock}) to (\ref{eq:selfdualcorrelator4intermsofbuildingblock}). All of the other remaining Green's functions obey the appropriate Ward identities (\ref{eq:wardidentitieslist}), and have the symmetry property $G^R_{AB}=G^R_{BA}$.

\bibliographystyle{JHEP}
\bibliography{HeatTransportDraft}

\end{document}